\documentclass[structabstract]{aa}  
\usepackage[pdftex]{graphicx}
\usepackage{amsmath}
\usepackage[multiple]{footmisc}
\usepackage{fixltx2e}
\usepackage{txfonts}
\begin{document}
\title{Star formation in the luminous YSO IRAS~18345-0641}

\author{Watson P. Varricatt
\inst{1}\thanks{e-mail:{w.varricatt@jach.hawaii.edu}}
\and
Holly S. Thomas
\inst{1}
\and
Chris J. Davis
\inst{2}
\and
Suzanne Ramsay
\inst{3}
\and
Malcolm J. Currie
\inst{1}
}
\institute{Joint Astronomy Centre,
	660 N. Aohoku Pl., Hilo, HI-96720, USA
\and
Astrophysics Research Institute, Liverpool John Moores University,
Birkenhead, Wirral, CH41 1LD, UK 
\and
ESO, Karl-Schwarzschild-Str. 2, 85748, Garching b. M{\"{u}}nchen, Germany
}

\date{Received January 10, 2013; accepted April 8, 2013}

 
\abstract
{} 
{We aim to understand the star formation associated with 
the luminous young stellar object (YSO) IRAS~18345-0641
and to address the complications arising from unresolved 
multiplicity in interpreting the observations of massive 
star-forming regions.}
{New infrared imaging data at sub-arcsec spatial resolution
are obtained for IRAS~18345-0641.
The new data are used along with mid- and far-IR imaging 
data, and CO ($J=3-2$) spectral line maps downloaded from 
archives to identify the YSO and study the properties 
of the outflow. Available radiative-transfer models are 
used to analyze the spectral energy distribution (SED) of 
the YSO.
}
{Previous tentative detection of an outflow in the 
H$_2$ (1-0) S1 line (2.122\,$\mu$m) is confirmed through 
new and deeper observations.  The outflow appears to be 
associated with a YSO discovered at infrared wavelengths.
At high angular resolution, we see that the YSO is probably 
a binary.  The CO~(3--2) lines also reveal a well defined 
outflow.  Nevertheless, the direction of the outflow deduced 
from the H$_2$ image does not agree with that mapped in CO.  
In addition, the age of the YSO obtained from the SED 
analysis is far lower than the dynamical time of the outflow.  
We conclude that this is probably caused by the contributions 
from a companion. High-angular-resolution observations at 
mid-IR through mm wavelengths are required to properly 
understand the complex picture of the star formation 
happening in this system, and generally in massive star 
forming regions, which are located at large distances from us.
}
{}


\maketitle
%

\section{Introduction}

It is now understood through recent studies that massive stars, 
at least up to late-O spectral types, form primarily through 
disk accretion and driving outflows (e.g. Arce et al. \cite{arce07}, 
Beuther et al. \cite{beuther02b}, Davis et al. \cite{davis04}, 
Varricatt et al. \cite{varricatt10}).  However, they are known 
to form in clusters in giant molecular clouds that are located 
mostly along the galactic plane at large distances from us.
The spatial resolution and wavelength coverage of most of the 
studies available are insufficient to probe their multiplicity,
which often lead to inaccuracies in the estimates of the 
properties of the YSOs, their disks and outflows. Detailed 
studies of individual sources are therefore necessary, which 
were often not feasible in surveys though which they were 
identified.  In this paper, we present the results of a 
multi-wavelength observational study of the luminous 
YSO IRAS~18345-0641 (hereafter IRAS~18345) and discuss the
complications associated with treating a YSO as single, where it
is possibly composed of two or more components.

IRAS~18345 was identified as a candidate ultra-compact H{\sc{ii}} 
region by van der Walt, Gaylard \& MacLeod (\cite{vanderwalt95}), 
who detected a 6.7\,GHz methanol maser from its vicinity. Highly 
variable and multi-component 6.7\,GHz Class-I methanol maser
emission has been later detected towards this source by several 
other investigators (Szymczak, Hrynek \& Kus \cite{szymczak02};
Szymczak et al. \cite{szymczak02}; Sridharan et al. \cite{sridharan02};
Beuther et al. \cite{beuther02c}; Pestalozzi, Minier \& Booth 
\cite{pestalozzi05}; Bartkiewicz et al. \cite{bartkiewicz09}).
Bronfman, Nyman \& May (\cite{bronfman96}) detected CS(2-1) 
emission from a dense core associated with IRAS~18345 at 
V$_{LSR}$=95.9~km~s$^{-1}$.
A distance of 9.5\,kpc has been estimated with the distance 
ambiguity solved (Sridharan et al. \cite{sridharan02}).  

H$_2$O maser emission is used as a tracer of the early 
stages of massive star formation, and is believed to be 
excited by jets (e.g. Felli, Palagi \& Tofani \cite{felli92}).  
This source hosts H$_2$O maser 
emission (Sridharan et al. \cite{sridharan02}; 
Beuther et al. \cite{beuther02c}), showing its youth. 
OH maser emission (1612~MHz), which is known to be a 
tracer of circumstellar disks, has also been detected 
towards IRAS~18345 (Edris, Fuller \& Cohen \cite{edris07}; 
V$_{peak}$=93.52~km~s$^{-1}$). 

Beuther et al. (\cite{beuther02b}) mapped a massive molecular 
outflow from this source in the $^{12}$CO ($J=2-1$) line.  The 
outflow was seen to be oriented NW-SE, with a projected length 
of 40\arcsec (1.84~pc) and a collimation factor {\it f}$_c$=1.5.  
C$^{18}$O~(2-1) observations by Thomas \& Fuller (\cite {thomas08}) 
detected a line that was best fit by a narrow and a broad Gaussian 
component at 95.65 and 95.79 km~s$^{-1}$ respectively indicating 
that they are likely to be associated with the same core or 
material; they suggest that the broad component is related to the 
outflow. Beuther et al. (\cite{beuther02a}) detected a possible 
outflow source at 1.2\,mm (integrated flux density=1.4~Jy) using  
MAMBO array and the IRAM 30-m telescope. This source was also 
detected at 1.1\,mm with an integrated flux of 0.92($\pm$0.12)~Jy
in the Bolocam Galactic Plane Survey conducted using the Caltech 
Submillimeter Observatory (Rosolowsky et al. \cite{rosolowsky10}), 
and at 450 and 850\,$\mu$m in the survey of high-mass protostellar
candidates using JCMT and SCUBA 
(Williams, Fuller \& Sridharan \cite{williams04}).

Varricatt et al. (\cite{varricatt10}) imaged this region in 
the near-IR $K$ band and in a narrow-band filter centred at 
the wavelength of the H$_2$ (1-0) S1 line at 2.122\,$\mu$m
as a part of their imaging survey of massive YSOs. They 
had a tentative detection of an outflow as an H$_2$ 
line emission knot (MHO~2211; see Figure~A20 of 
Varricatt et al). They detected two extremely red near-IR 
sources (labelled
`A' [$\alpha$ = 18:37:17.00, $\delta$ = -06:38:24.5] and
`B' [$\alpha$ = 18:37:16.91, $\delta$ = -06:38:30.7]{\footnote{All coordinates
given in this paper are in J2000}}{\footnote{The RAs were erroneously reported by them as
$\alpha$ = 18:37:7.00 and 18:37:6.91 for `A' and
`B' respectively.}}). Source `B' detected by them was very 
close to the IRAS source, the 1.2-mm continuum peak of 
Beuther et al. ({\cite{beuther02a}) and the CH$_3$OH masers 
detected by Beuther et al. ({\cite{beuther02c}).

IRAS~18345 is likely to be in a pre-UCH{\sc{ii}} 
phase. Radio emission is very weak towards this source.
Hofner et al. (\cite{hofner11}) detected a faint 
(210\,$\mu$Jy) unresolved 1.3-cm radio continuum source 
located at $\alpha$=18:37:16.91 $\delta$=-06:38:30.4.
They propose that the radio
emission was from an ionized jet driving the massive 
CO outflow detected by Beuther et al. ({\cite{beuther02b}).
(Sridharan et al. (\cite{sridharan02}) reported a 27-mJy
3.6-cm source towards IRAS~18345.  However, 
Hofner et al. (\cite{hofner11}) pointed out that this emission
was from an unrelated source located 3\arcmin SE of the IRAS 
source). 

\section{Observations and data reduction}

\subsection{UKIRT data}
\label{ukirtdata}

\subsubsection{WFCAM near-IR data}

We observed this region on multiple epochs using the United
Kingdom Infrared Telescope (UKIRT), Hawaii, and the UKIRT 
Wide Field Camera (WFCAM; Casali et al. \cite{casali07}) 
during the UKIDSS backup time.  Observations were 
performed using the near-IR {\em J, H} and {\em K} MKO consortium 
filters and a narrow band filter centered at the wavelength 
of the H$_2$ (1-0) S1 line at 2.1218~$\mu$m.  WFCAM employs
four 2048$\times$2048 HgCdTe HawaiiII-2RG arrays; for the
observations presented here, we used the data from only
one of the four arrays in which the object was located. 
At a pixel scale of 0.4\arcsec~pixel$^{-1}$, each array
has a field of view of 13.5\arcmin$\times$13.5\arcmin.
Observations were carried out by dithering the object
to 9 positions separated by a few arcseconds in {\em J, H}
and {\em K} bands, and to 5 positions in H$_2$, and using
a 2$\times$2 microstep, resulting in a pixel scale
of 0.2\arcsec~pixel$^{-1}$. Table \ref{tab:NIR-obslog}
gives a log of the WFCAM observations performed.

\begin{table}
\caption{Log of WFCAM observations}     
\label{tab:NIR-obslog}                  
\centering                              
\begin{tabular}{llllll}                 
\hline\hline                                    \\[-2mm]
UTDate          &Filters        &Exp. 	&Total int.          &FWHM           \\[1mm]
({\small{yyyymmdd}})&used       &time (s) &time (s)               &(arcsec)       \\
\hline  \\[-2mm]
20120401        &$K$            &5,1	&180,72       		&0.82,0.84	\\[1MM]
20120407        &$J, H$         &10,5	&180,360             	&1.11,0.99	\\[1MM]
20120426        &$J$            &10     &360                    &0.66           \\[1mm]
20120501        &$H_2$          &40     &800,800                &0.95,0.83	\\[1mm]
20120522        &$H_2, K, K$    &40,5,1	&800,180,72   		&1.26,1.18,1.16 \\[1mm]
\hline
\end{tabular}
\end{table}

The data reduction, and archival and distribution were carried
out by the Cambridge Astronomical Survey Unit (CASU) and the
Wide Field Astronomy Unit (WFAU) respectively. Since the 
photometric calibrations are performed using point sources 
present in all dithered frames and using their 2MASS magnitudes, 
the photometric quality is good even in the presence of clouds.
The photometric system and calibration are described in 
Hewett et al. (\cite{hewett06}) and Hodgkin et al. 
(\cite{hodgkin09}) respectively. The pipeline processing and
science archive are described in Irwin et al. (\cite{irwin04})
and Hambly et al (\cite{hambly08}). The left panel of 
Figure~\ref{wfcamJHKH2} shows a {\em JHK} colour composite image 
created from the WFCAM data in a 1.5$\arcmin\times$1.5$\arcmin$ 
field centred on IRAS~18345. Sources `A' and `B' are labelled 
on the figure.  Both sources are detected in {\em K}. This region 
was also observed in the $JHK$ bands using WFCAM as a part of 
the UKIRT Infrared Deep Sky Survey (UKIDSS; Lawrence et al. 
\cite{lawrence07}). We downloaded the {\em JHK} images (obtained 
for the Galactic Plane Survey, GPS) and merged catalogs from 
the UKIDSS GPS DR6+.

Source `A' is saturated in our images with 5~sec exposure, so its 
{\em K} magnitude is derived from a mosaic obtained with per frame 
exposure time of 1 sec on 20120522~UT. It is detected well in {\em H} 
band.  We detect a very faint source at the location of 
`A' in the $J$ band image obtained on 20120426~UT, when the seeing
was good.  It has a faint neighbour $\sim$0.8\arcsec NE.  Hence 
its magnitudes are measured using an aperture of radius 0.4\arcsec
and applying aperture corrections.  `B' is detected well in $K$
as a point source embedded in nebulosity.
Its estimated magnitudes are affected by the presence of nebulosity,
with the {\em K}-magnitudes in a 3\arcsec-diameter aperture brighter
than that in a 2\arcsec-diameter aperture by ~0.21 mag. `B' is
not detected in {\em J} and only the nebulosity associated with it 
is detected in {\em H}.  Table \ref{tab:wfcamJHK} shows the magnitudes
of the two sources measured on different epochs.  

\begin{figure*}
\centering
\includegraphics[width=18.0cm]{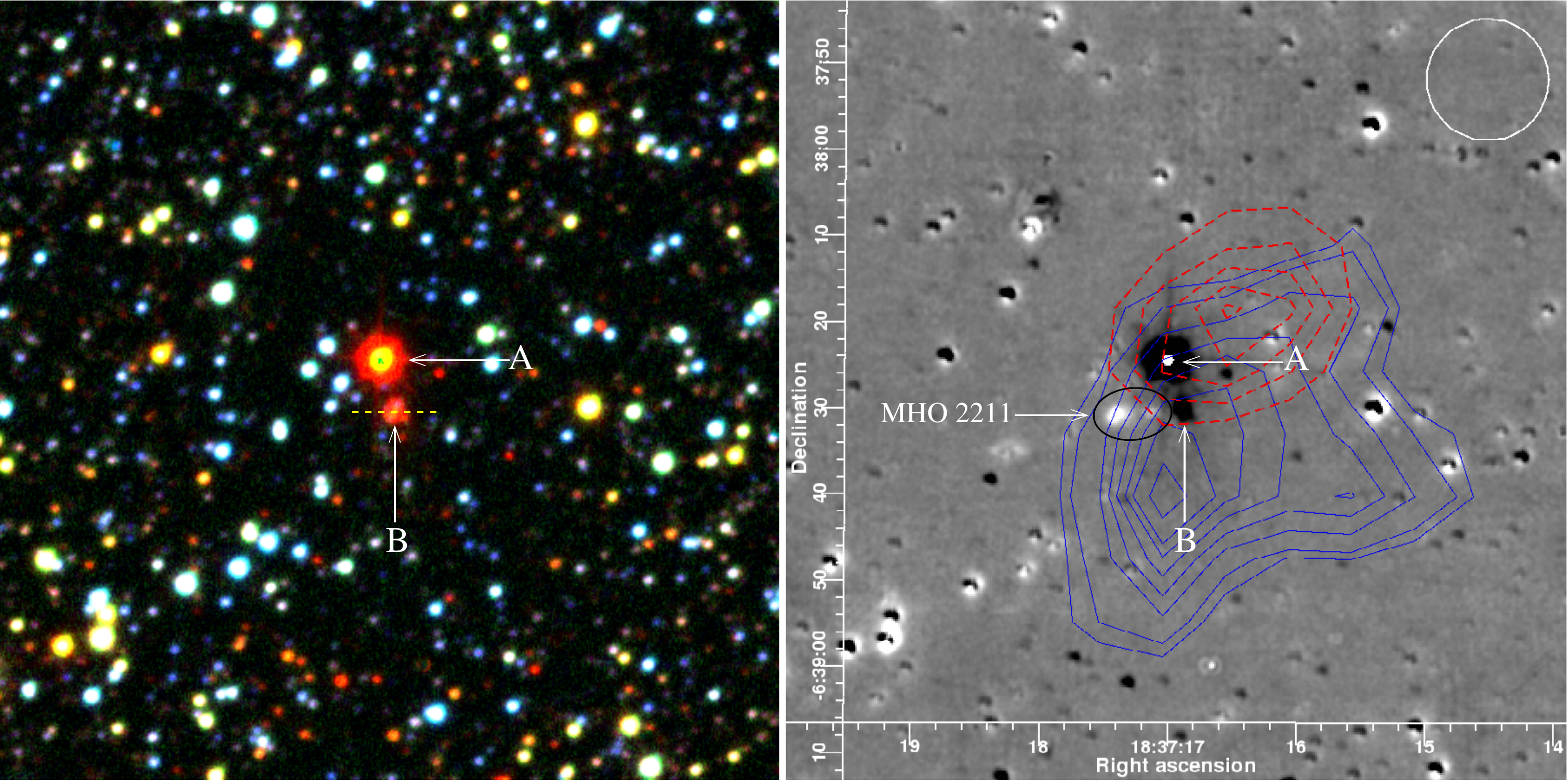}
\caption{The left panel shows WFCAM $JHK$ colour-composite 
image ($J$-blue, $H$-green, $K$-red), and the right panel 
shows the continuum-subtracted H$_2$ image, smoothed with 
a Gaussian of FWHM=2 pix, and averaged over all our observations.  
The images are constructed from our new WFCAM data, and cover 
a filed of view of 1.5\arcmin$\times$1.5\arcmin. The 
dashed yellow line on the $JHK$ image shows the location 
and direction of the major axis (length $\times$50) of the 
elliptical distribution of the 6.7\,GHz methanol maser spots 
detected by Bartkiewicz et al. (\cite{bartkiewicz09}).  The 
blue (continuous) and red (dashed) contours overlaid on the
H$_2$ image show the deconvolved blue- and red-shifted 
lobes of the outflow detected in CO (J=3-2) (see \S\ref{COdata}); 
the blue contours are at 3, 5, 10, 15, 20, 25, 30 and 
40\,$\sigma$, and the red contours are at 3, 5, 7, 9, and 
11\,$\sigma$ above the mean background.  The H$_2$ line 
emission features are enclosed in ellipse and labelled 
``MHO 2211''. The white circle on the top-right corner 
shows the JCMT beam size (14.6\arcsec at 345\,GHz).  
}
\label{wfcamJHKH2}%
\end{figure*}

The H$_2$ images were continuum subtracted using scaled
$K$ band images. The background emission was fitted and
removed from the H$_2$ and $K$ band mosaics. The seeing
obtained in our {\em K} band mosaics was better than in H$_2$.
For any H$_2$ mosaic, the {\em K} band mosaic with the closest
seeing was used for continuum subtraction, after
smoothing the {\em K} band mosaic with a suitable Gaussian to
match FWHM in the H$_2$ mosaic.
The average of the ratio of counts {\em K}/H$_2$ was evaluated
for a few isolated point sources present in both {\em K} and 
H$_2$ mosaics. The background-subtracted $K$ mosaics were
scaled by these ratios and were subtracted from the
background-subtracted H$_2$ mosaics to perform the continuum
subtraction. The three H$_2$ mosaics were continuum subtracted, 
smoothed with a Gaussian of FWHM=2 pixels to enhance the 
appearance of the faint line emission features and averaged. 
The right panel of Figure~\ref{wfcamJHKH2} shows the
continuum-subtracted H$_2$ image of the same region as
in the {\em JHK} colour composite in the left.

\begin{table} 
\caption{Near-IR magnitudes of sources `A
and `B'.  A blank space is given when the observations
were not done in a particular filter on any given UTDate.}
\label{tab:wfcamJHK}      
\centering               
\begin{tabular}{llllll}   
\hline\hline                                                            \\[-2mm]
UTdate	  &\multicolumn{3}{c}{Magnitudes$^{\mathrm{a}}$}		\\
YYYYmmdd &$J$	&$H$     	 &$K$                   		\\
\hline                                                                  \\[-2mm]
\multicolumn{4}{l}{Source `A'; $\alpha$=18:37:17.016, $\delta$=-6:38:24.36}\\
20070826$^{\mathrm{b}}$ &not resolved   &12.82 (0.01)  &Saturated      \\
20120407 &not detected  &12.81 (0.01)   &				\\
20120426 &20.14 (0.15)	&		&				\\
20120522 &		&		&8.82 (0.01) 			\\
\hline\\[-2mm]
\multicolumn{4}{l}{Source `B'; $\alpha$=18:37:16.896, $\delta$=-6:38:30.48}\\
20031009$^{\mathrm{c}}$ &               &		&12.69 (0.01)  \\
20070826$^{\mathrm{b}}$ &               &15.934 (0.016) &12.45 (0.01)  \\
20120401 &              &               &12.82 (0.01)  		\\
20120407 &not detected	&16.17 (0.05)	&				\\
20120522 &not detected	&		&12.53 (0.01)  		\\
\hline
\end{tabular}
\begin{list}{}{}
\item[$^{\mathrm{a}}$The magnitudes] are in the UKIRT photometric
system, measured using an aperture of diameter 2\arcsec and applying 
aperture corrections. The values given in parenthesis are the 
1-$\sigma$ internal errors. $^{\mathrm{b}}$From UKIDSS GPS DR6+.  
$^{\mathrm{c}}$From the images presented in Varricatt et al. 
(\cite{varricatt10}) observed using UKIRT and UFTI; the data of 
all other nights were obtained using UKIRT and WFCAM.
\end{list}
\end{table}

\subsubsection{IRCAM3 imaging in $L'$ and $M'$ bands}

$L'$ and $M'$ imaging observations of IRAS~18345 were performed 
using UKIRT and the facility 1--5-$\mu$m imager 
IRCAM3{\footnote{http://www.jach.hawaii.edu/UKIRT/instruments/ircam/ircam3.html}}.
IRCAM3 uses a 256~$\times$~256 InSb array and gives a total field 
of view of 20.8\arcsec$\times$20.8\arcsec with an image scale of 
0.081\arcsec~pix$^{-1}$ at the Cassegrain focal plane of UKIRT.  
The $L'$ and $M'$ observations were performed on 20000715 and 
20000716~UT respectively.  Individual exposure times of 0.2~sec
and 0.12~sec with 100 coadds were used in $L'$ and $M'$ respectively.
Observations were performed by offsetting the field to 4 positions 
on the array, separated by 15\arcsec each in RA and Dec. This 
4-point jitter pattern was repeated thrice. The adjacent frames were 
subtracted from each other and are flat fielded.  This resulted in 
a mosaic with two positive and two negative beams for the target. 
Two of the four offset positions were outside the array and in one
of the remaining two, source `A' was outside the array;  hence, 
the total exposure time for `A' and `B' were 60 and 120~sec 
respectively in $L'$ and 36 and 72~sec respectively in $M'$.  

The sky conditions were photometric on both nights.  The excellent 
seeing on 20000716 ($\sim$0.34\arcsec) during the $M'$ observations 
enabled us to resolve source `B' into two components `B1' and `B2' 
for the first time.  These two components are at a separation of 
$\sim$0.45\arcsec in the SE-NW direction, with `B1' located SE of 
`B2'.  Figure~\ref{ircamM} shows the $M'$ image of the central 
region of IRAS~18345 with the sources `A' and `B' detected.  The 
contours generated from the $K$-band image are overlaid on 
Figure~\ref{ircamM}.  The insets on the right show a 
1.5\arcsec~$\times$~1.5\arcsec field containing source `B'
in $M'$. `B1' and `B2' are labelled on the figure. 

When compared with that in $M'$, the seeing was poor and variable 
during the $L'$ observations (FWHM$\sim$0.61\arcsec in the mosaic), 
hence `B' is not resolved in the final mosaic in $L'$.  However, 
the seeing was better (FWHM$\sim$0.38\arcsec) for two offset 
positions (40-s exposure) of the $L'$ observations; an $L'$ mosaic
in a 1.5\arcsec$\times$1.5\arcsec field around `B' constructed
using those two frames is shown in the bottom-left inset in 
Figure~\ref{ircamM}.  Photometric standards were observed in 
$L'$ and $M'$ for flux calibration.  The magnitudes were derived 
using an aperture of diameter 4.07\arcsec (50 pixels) and 
applying extinction corrections for the differences in the 
airmass between the object and the standard stars. The $M'$-band 
photometry of  `B1' and `B2' were derived using a 
0.326\arcsec(4 pixels)-diameter aperture and applying aperture 
corrections (derived from source `A' in the $M'$-band
mosaic, and from standard stars for the $L'$ frames). 
Source `B' detected in our $K$-band image is centred on
`B1', with a nebulous extension in the direction of `B2'.
Astrometric calibration of the $L'$ and $M'$  images were 
performed using the coordinates of `A' and `B1' measured 
from the $K$-band image. The coordinates and magnitudes of 
the sources detected are given in Table~\ref{tab:LM12p5}.

\begin{table*}
\caption{IRCAM3 imaging in $L'$ and $M'$ bands and Michelle imaging at 12.5\,$\mu$m}
\label{tab:LM12p5}      
\centering               
\begin{tabular}{llllll}   
\hline\hline                                            		\\[-2mm]
Source 	&RA$^{\mathrm{a}}$     	&Dec$^{\mathrm{a}}$    &\multicolumn{2}{c}{IRCAM3 magnitudes$^{\mathrm{b}}$}	&Michelle flux (Jy)$^{\mathrm{b}}$	\\
identification &    		&   	&$L'$ &$M'$					&12.5\,$\mu$m	\\
\hline                                                  				\\[-2mm]
A       &18:37:17.005   &-06:38:24.5	&5.15 (0.01) 	&4.59 (0.04)	&4.53 (0.18)		\\
B1+B2	&		&	   	&8.53 (0.02) 	&6.71 (0.05)	&1.56 (0.10)	\\
B1      &18:37:16.917   &-06:38:30.77   	&9.65 (0.09)	&7.21 (0.06)	&		\\
B2      &18:37:16.897   &-06:38:30.43   	&10.12 (0.09)	&7.98 (0.06) 	&		\\
\hline
\end{tabular}
\begin{list}{}{}
\item$^{\mathrm{a}}$From the IRCAM images.  $^{\mathrm{a}}$The values given in parenthesis are the 1-$\sigma$ errors - in the zero points for IRCAM3, and in the flux measured in the 4 beams for MICHELLE. 
\end{list}
\end{table*}

\begin{figure}
\centering
\includegraphics[height=8.85cm, width=8.9cm]{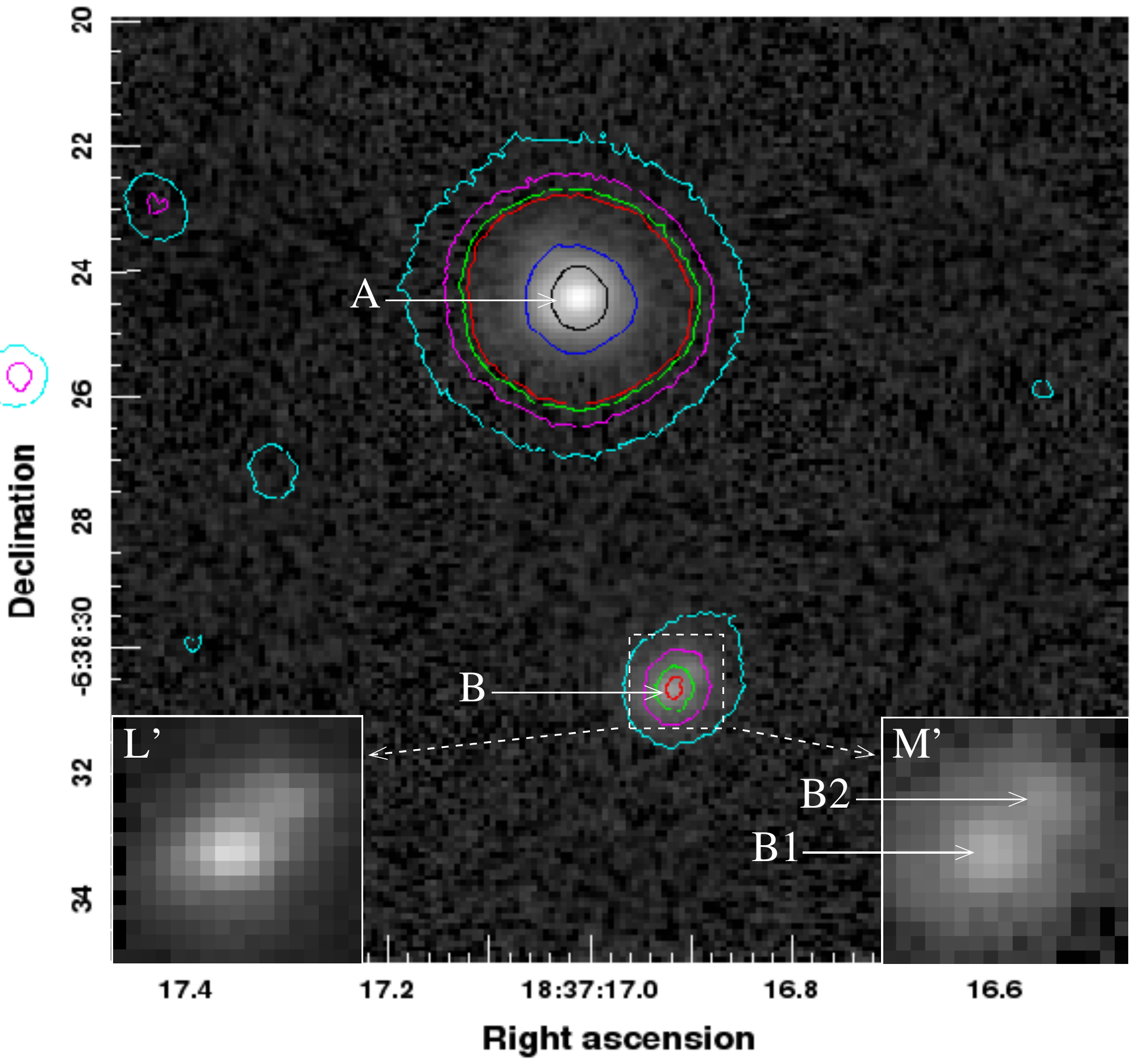}
\caption{$M'$-band image of the central 15\arcsec$\times$15\arcsec field around IRAS~18345
Contours from the UFTI $K$-band image of Varricatt et al. (\cite{varricatt10}) are 
overlaid. The bottom-left and bottom-right insets show 1.5\arcsec$\times$1.5\arcsec fields 
surrounding source `B' in $L'$ and $M'$ respectively.
}
\label{ircamM}%
\end{figure}

\subsubsection{Michelle imaging at 12.5\,$\mu$m}

We observed IRAS~18345 using using UKIRT and
Michelle{\footnote{http://www.jach.hawaii.edu/UKIRT/instruments/michelle/\\michelle.html}}
on 20040329~UT. Michelle (Glasse et al. \cite{glasse97})
is a mid-infrared imager/spectrometer using an SBRC Si:As
320x240-pixel array and operating in the 8--25\,$\mu$m
wavelength regime.  It has a field of view of
67.2\arcsec$\times$50.4\arcsec with an image scale of
0.21\arcsec/pix at the Cassegrain focal plain of UKIRT.

The Michelle observations were obtained using a filter
centered at 12.5\,$\mu$m with 9\% passband.  The sky 
conditions were photometric. BS~6705 and BS~7525 were 
used as standard stars.  The observations were performed 
by chopping the secondary in the N-S direction and 
nodding the telescope in the E-W by 15\arcsec each 
(peak-peak) to detect faint sources in the the presence 
of strong background emission.  Data reduction was 
performed using the UKIRT pipeline {\sc{orac-dr}} 
(Cavanagh et al. \cite{cavanagh08})  and 
Starlink {\sc{kappa}} (Currie et al. \cite{currie08}).  
The resulting mosaic has four images 
of the sources detected (2 positive and 2 negative 
beams). The final image was constructed after negating 
the negative beams and combining all four beams.   
Figure~\ref{mich12p5} shows a 15\arcsec$\times$15\arcsec 
field of the mosaic surrounding IRAS~18345. An average 
FWHM of 0.98\arcsec was measured from the sources 
detected in the 12.5\,$\mu$m image. The astrometric 
corrections were applied by adopting the position of 
source `A' from Varricatt et al. (\cite{varricatt10}). 
The coordinates derived for the fainter 12.5-$\mu$m 
source agrees well with that of the near-IR source 
`B' detected by Varricatt et al. (\cite{varricatt10}).  
The fluxes measured for `A' and `B' are given in 
Table \ref{tab:LM12p5}. At this spatial resolution, 
the two components of `B' are not resolved.

\begin{figure}
\centering
\includegraphics[height=8.85cm, width=8.85cm]{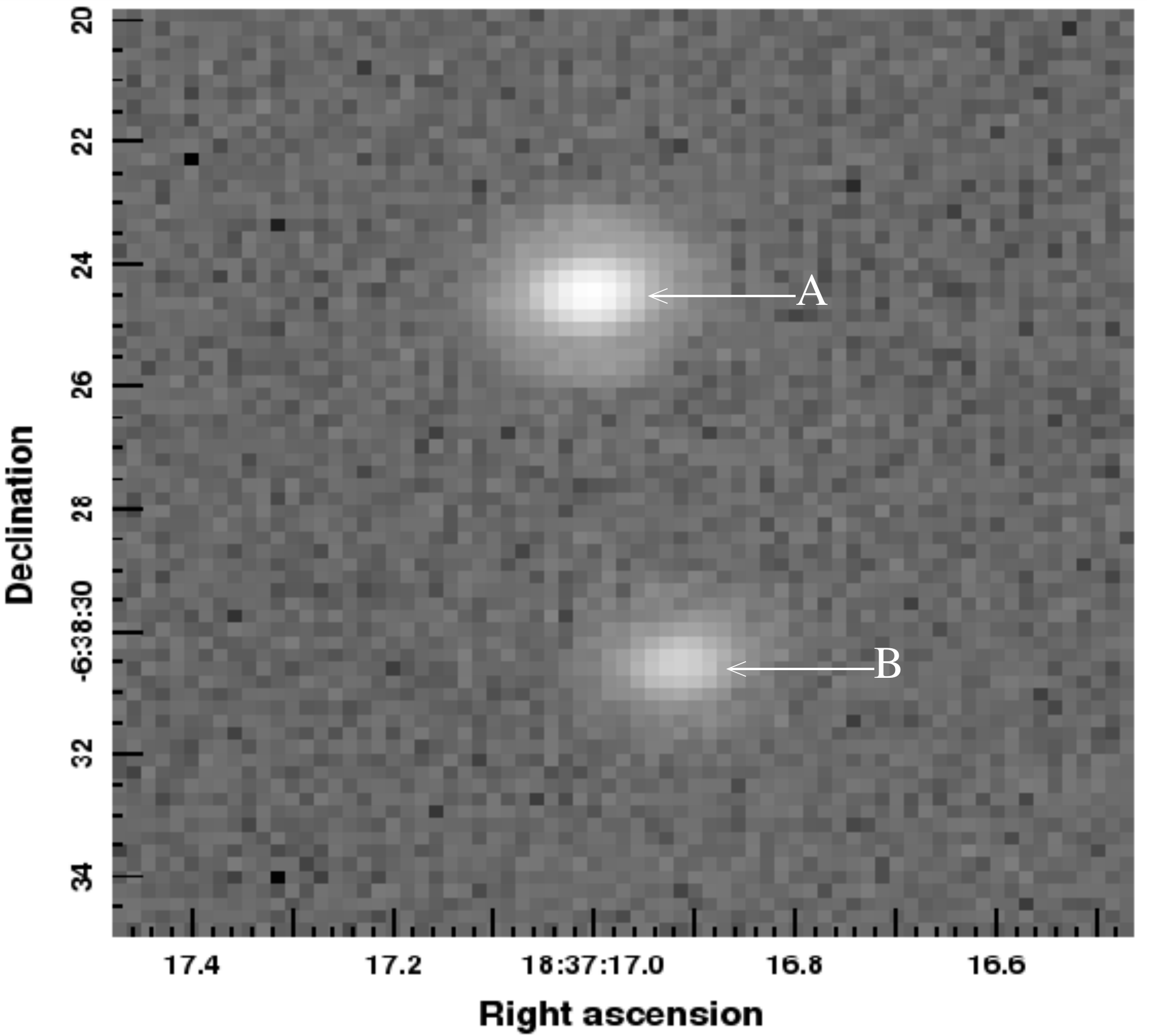}
\caption{Michelle 12.5-$\mu$m image of IRAS~18345 covering
a field of 15\arcsec$\times$15\arcsec.
}
\label{mich12p5}%
\end{figure}

\subsection {Methanol observations}

The $J$=5--4 (241.80\,GHz) and $J$=7--6 (338.53\,GHz) transitions
of methanol were observed using the 15-m James Clerk Maxwell 
Telescope (JCMT), Hawaii, in July and August 2004. The receiver 
RxA (211--279 GHz) was used for the
$J=5-4$ transition and receiver RxB3 (315--373\,GHz) was
used for $J$=7--6. In both cases the back end was the DAS
(Digital Autocorrelator Spectrometer) with a bandwidth and
spectral resolution of 250\,MHz and 0.16\,MHz respectively for
$J$=5--4, and 500\,MHz and 0.625\,MHz  respectively for $J$=7--6.
Pointing at the JCMT was checked every hour and had an
rms $<$2.5$''$. The weather was good during the observations,
with opacities in the range 0.13$<\tau_{225}<$0.16. The data
were reduced using the data reduction package {\sc{specx}} 
(Prestage et al. \cite{prestage00}). Figure~\ref{CH3OH} 
shows the methanol spectra.

CH$_3$OH, is an organic molecule that originates primarily 
on the surface of grains, usually seen towards hot cores.
Organic molecules such as these appear in the gas phase once 
sufficient ices have been evaporated from the grain surfaces. 
This implies gas temperatures in excess of 100\,K 
(Rodgers \& Charnley \cite{rodgers03}). These conditions are 
achieved either through the localized heating of the region 
immediately surrounding the young star, or in the shocked 
regions produced by a powerful outflow. 

Table \ref{tab:mm} shows the rest frequencies, observed 
radial velocities, intensities and line widths of the prominent 
methanol lines detected in our spectra. The line identifications 
are adopted from Anderson, De Lucia \& Herbst (\cite{anderson90}) and 
Maret et al. (\cite{maret05}). The lines are mostly centred 
at the $\upsilon$\textsubscript{LSR} of the core, and have
large width, suggesting that they originate in a warm region
surrounding the core.

\begin{figure}
\centering
\includegraphics[width=9cm]{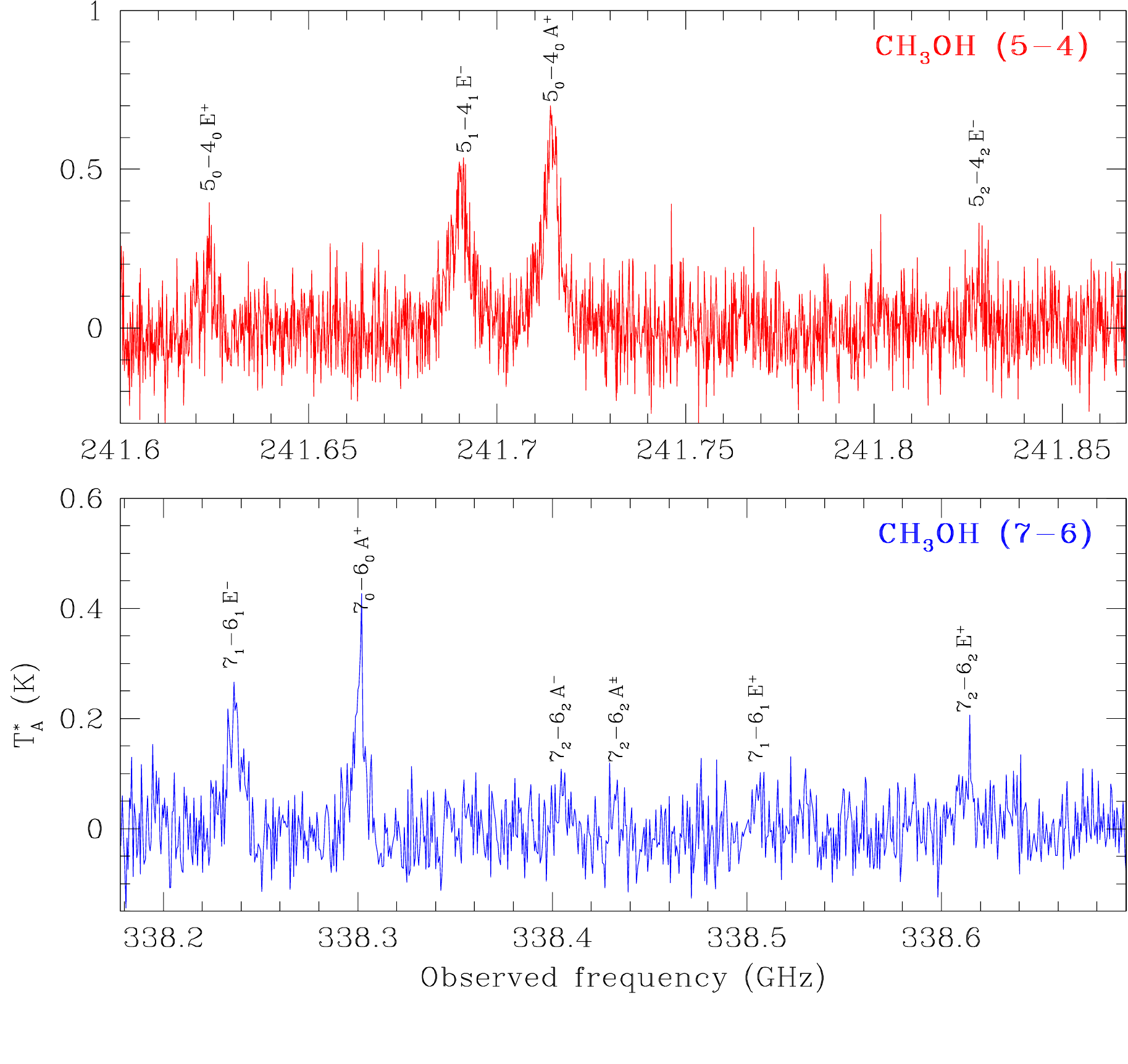}
\caption{CH$_3$OH~($J$=7--6; 338.53\,GHz) and ($J$=5--4; 241.80\,GHz) spectra observed using JCMT.
}
\label{CH3OH}
\end{figure}

\begin{table}  
\setlength{\tabcolsep}{0.45em}
\caption{\bf Methanol lines detected}
\label{tab:mm}      
\centering                      
\begin{tabular}{lllllll}       
\hline\hline                                    \\[-2mm]
Line ID         &Rest Freq.	&{E\textsubscript{up}}&Radial	&Width       	&Amp.	        &$\int$T$_A^*$     \\
		&(GHz)		&(K)     &Vel.	&(km/s)		&T$_A^*$(K)     &(K~km/s) \\
\hline                                          \\[-2mm]

\multicolumn{2}{l}{CH$_3$OH (7--6)}               &	&               &               &               \\
7$_1$--6$_1$ E$^-$ &338.344605 	&69.61   &95.4	&7.25		&0.16           &1.21           \\
7$_0$--6$_0$ A$^+$ &338.408718	&64.98   &95.4	&5.33 		&0.28           &1.60           \\
7$_2$--6$_2$ E$^+$ &338.721694 	&86.31   &95.5	&9.22 		&0.06           &0.61           \\
\hline\\[-3mm]
\multicolumn{2}{l}{CH$_3$OH (5--4)}               &	&               &               &               \\
5$_0$--4$_0$ E$^+$ &241.700168  &46.99   &95.2	&8.03		&0.13           &0.97           \\
5$_1$--4$_1$ E$^-$ &241.767247  &39.45   &95.2	&8.88		&0.37           &3.54           \\
5$_0$--4$_0$ A$^+$ &241.791367	&34.82   &95.4	&7.16   	&0.53           &4.35           \\
5$_2$--4$_2$ E$^-$ &241.904643  &59.78   &95.2	&5.11		&0.11           &0.52           \\
\hline
\end{tabular}
\end{table}

\subsection{Archival data}

IRAS18345 was detected well in the sky surveys conducted
using the Wide-field Infrared Survey Explorer (WISE; 
Wright et al. \cite{wright10}) and the AKARI satellite 
(Murakami et al., \cite{murakami07}). The WISE mission 
surveyed the sky in four bands - $W$1 (3.4\,$\mu$m), 
$W$2 (4.6\,$\mu$m), $W$3 (12\,$\mu$m) and $W$4 (22\,$\mu$m). 
At the 6.1$\arcsec$--12.0$\arcsec$ spatial resolution of WISE, 
sources `A' and `B' are not resolved; however, the 
excellent positional accuracy (0.15$\arcsec$) of the WISE data
enables the identification of the source.  The location of the
WISE source is in between `A' and `B'; at the shorter wavelengths 
the centroid is closer to `A', while at 22\,$\mu$m, {\bf it is} 
located very close to `B' (see Figure \ref{IRAC123}) 
confirming that `B' is the leading contributor at longer 
wavelengths. At the FWHM of $\sim$5.7\arcsec at 9 and 
18\,$\mu$m (Ishihara et al. \cite{ishihara10}), and 30--40\arcsec 
at 65, 90, 140 and 160\,$\mu$m (Doi et al. \cite{doi09}) of the 
AKARI data, sources `A' and `B' are not resolved.

This region was observed in the {\it Spitzer} GLIMPSE II survey 
using the Infrared Array Camera (IRAC; Fazio et al. \cite{fazio04}) 
in bands 1--4, centered at 3.6, 4.5, 5.8 and 8.0\,$\mu$m 
respectively. {\it Spitzer} also observed this region using the 
Multiband Imaging Photometer for {\it Spitzer} (MIPS; Rieke et al. 
\cite{rieke04}) at 24\,$\mu$m and 71\,$\mu$m.  The reduced 
{\it Spitzer} images and point source photometry were downloaded 
from the NASA/IPAC Infrared Science Archive. Figure~\ref{IRAC123} 
shows a colour composite image of a 1$\arcmin\times$1$\arcmin$ 
region constructed from the {\it Spitzer} images at 3.6, 4.5 and 
5.8\,$\mu$m.  The {\it Spitzer} catalog gives magnitudes for source 
`A' in Bands 1--3 only and for source `B' in Bands 1, 3 and 4 only.  
We performed aperture photometry on the {\it Spitzer} images using 
a 3.6$\arcsec$-diameter aperture in Bands 1-3 and using a 
4.8$\arcsec$-diameter aperture in Band 4.  Average zero points were 
derived using a few isolated point sources in the field.  We derive 
magnitudes 5.44$\pm$0.05, 4.23$\pm$0.05, 3.71$\pm$0.05 and 
3.35$\pm$0.04 for source `A', and 9.13$\pm$0.05, 6.68$\pm$0.05, 
5.85$\pm$0.05 and 4.61$\pm$0.04 for source `B' in bands 1, 2, 3 and 
4 respectively.  The errors given are 1-$\sigma$ in the estimate 
of the zero points. IRAS~18345 appears to be close to saturation 
in IRAC bands 1 and 3. Hence we do not use the magnitudes derived 
in these bands in the SED analysis.

IRAS~18345 is saturated in the {\it Spitzer}-MIPS 24-$\mu$m image, 
but detected well at 71\,$\mu$m.  At an FWHM of 21\arcsec measured 
on source, `A' and `B' are not resolved at 71\,$\mu$m.  Using MOPEX, 
we estimate a flux of 71.8\,Jy for `A' and `B' together at 70\,$\mu$m.
The locations of the {\it Spitzer} source detected at 24\,$\mu$m and
70\,$\mu$m are also plotted on Figure~\ref{IRAC123}.

\begin{figure}
\centering
\includegraphics[width=9cm]{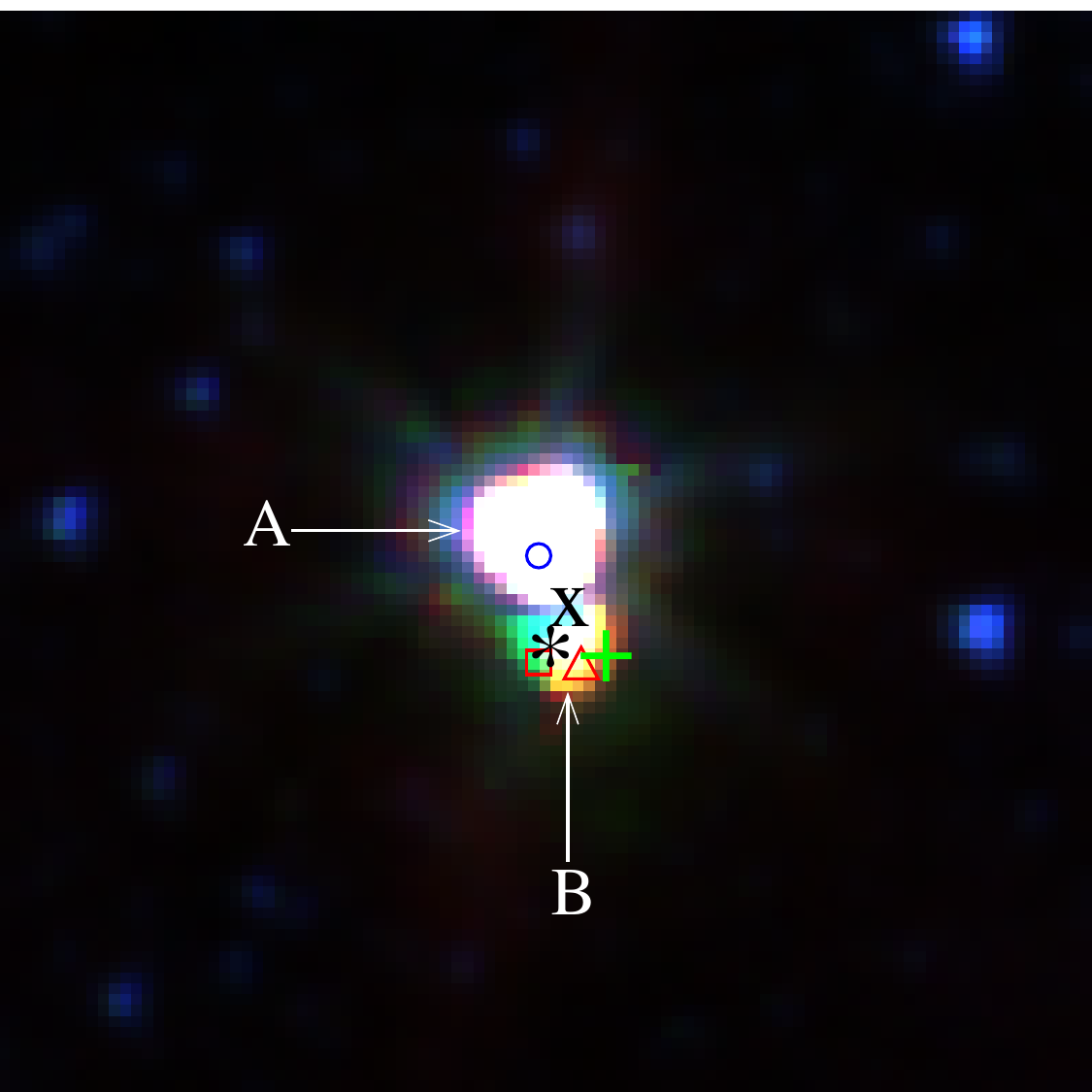}
\caption{{\it Spitzer}-IRAC colour composite image
(blue-band1, green-band2, red-band-3)
a 1\arcmin$\times$1\arcmin field centred on IRAS~18345.  The
symbols $\circ$, $\ast$, X, +, $\square$ and $\Delta$ show the locations of the
source detected by  
WISE at 3.4\,$\mu$m and 22\,$\mu$m, {\it Spitzer}-MIPS at 24\,$\mu$m and 70\,$\mu$m,
Beuther et al. (\cite{beuther02a}) at 1.2\,mm using IRAM 30-m telescope amd MAMBO, and 
and Williams et al. (\cite{williams04})
at 850\,$\mu$m using JCMT and SCUBA.
respectively. }
\label{IRAC123}
\end{figure}

\subsubsection {CO\,($J$=3--2) data}
\label{COdata}

CO observations were performed using the JCMT in 2008.
The $^{12}$CO\,(3--2) (345.796\,GHz) maps were observed on
20080421~UT, while the  $^{13}$CO\,(3--2) (330.588\,GHz) and 
C$^{18}$O~(3--2) (329.331\,GHz) maps were simultaneously observed
on 20080707~UT. All observations were performed in 
position-switched raster-scan mode and using half array spacing
with HARP (Buckle et al. \cite{buckle09}) as the front end.
The ACSIS autocorrelator was used as the back end. ACSIS was setup 
in dual sub-band mode for the  $^{13}$CO/C$^{18}$O observations 
giving a bandwidth of 250\,MHz and a resolution of 0.06\,MHz 
(0.055\,km~s$^{-1}$) for each line. The $^{12}$CO data were 
taken in single sub-band mode with a bandwidth of 1000\,MHz and 
a resolution of 0.488\,MHz (0.42\,km~s$^{-1}$). The atmospheric 
opacity at 225\,GHz, measured with the CSO dipper, was 0.26 for 
$^{12}$CO and 0.07--0.08 for $^{13}$CO/C$^{18}$O.  The telescope
pointing was checked and corrected before the observations.
The pointing accuracy on source is expected to be better than 
2\arcsec.

The CO data were taken from the JCMT Science Archive where 
they had been reduced using the {\sc orac-dr} pipeline. The 
pipeline first performs a quality-assurance check on the 
time-series data before trimming and de-spiking. It then applies 
an iterative baseline removal routine before creating the final 
group files. 

Figure~\ref{12CO-cube} shows a 3.3\arcmin$\times$3.3\arcmin 
field of the $^{12}$CO spectral cube, which shows well defined 
outflow. Figure~\ref{13CO-C18O-cube} shows the same field in 
$^{13}$CO and C$^{18}$O lines. $^{13}$CO\,(3--2) is also detected 
in outflow whereas in C$^{18}$O, we mostly detect the emission 
from the central core. Figure~\ref{COspectra} shows 
plots of the integrated spectra of $^{12}$CO, $^{13}$CO and 
C$^{18}$O in a 1\arcmin$\times$1\arcmin field around IRAS~18345.  
The dotted-dashed vertical line shows the rest velocity of the 
cloud. The emission at $\sim$110~km~s$^{-1}$ (Figures~\ref{12CO-cube},
\ref{13CO-C18O-cube} and \ref{COspectra}) is probably not related 
to the outflow.


\begin{figure*}
\centering
\includegraphics[width=15.0cm]{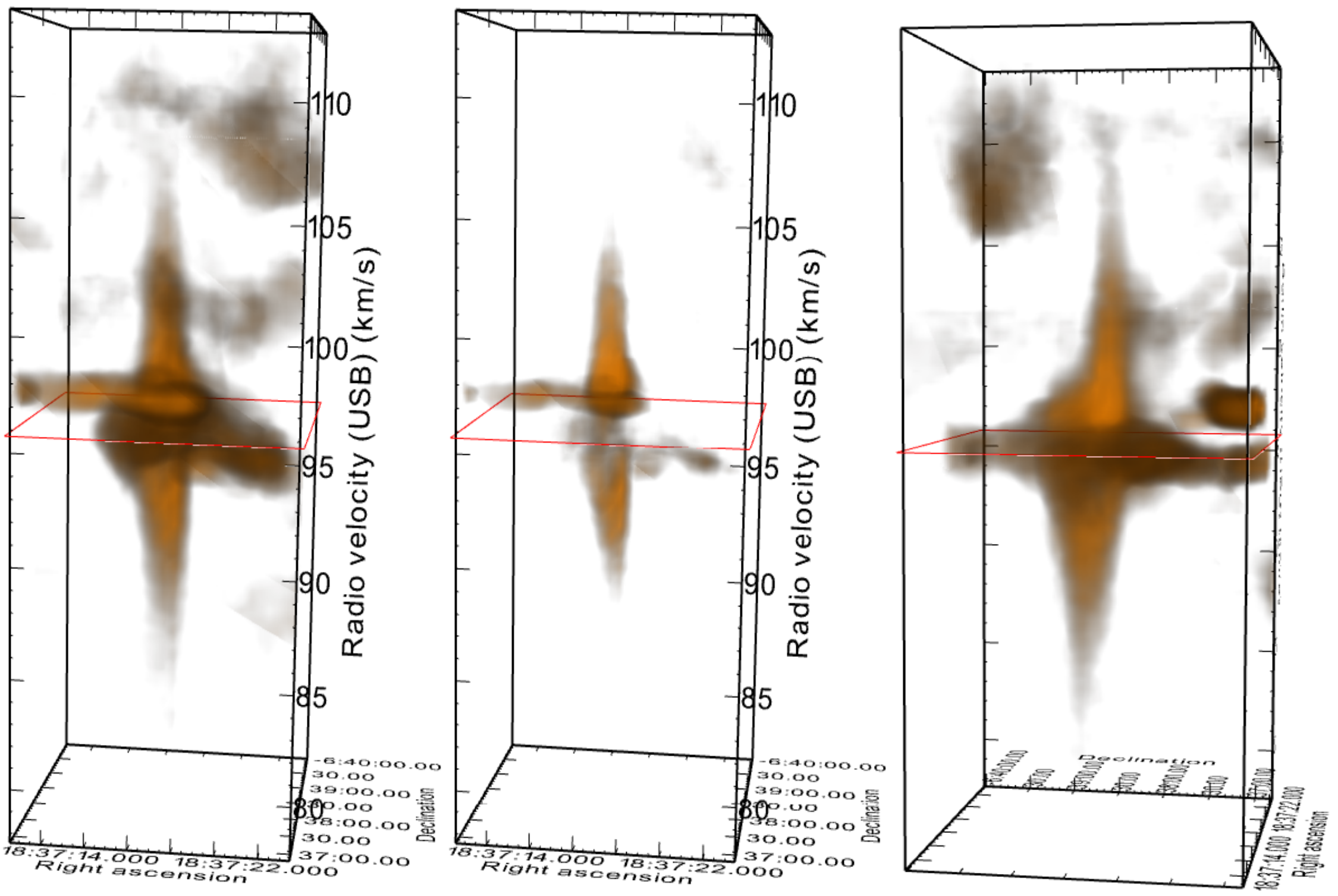}
\caption{$^{12}$CO~(J=3-2) spectral cubes covering
a 3.3\arcmin$\times$3.3\arcmin field surrounding IRAS~18345 
in a velocity range 77.6~km~s$^{-1}$ to 113.5~km~s$^{-1}$ 
centred on the CO-line rest frequency.  The left and right 
panels show the spectral cube in two different orientations.
The central panel shows the same region as in the left panel,
but with the faint emission filtered out to show the outflow
more clearly. The red square shows the plane of the 
$\upsilon$\textsubscript{LSR} of the core. The contours 
representing the blue- and red-shifted outflow lobes 
projected on to the sky plane are shown in Fig. \ref{wfcamJHKH2}.}
\label{12CO-cube}%
\end{figure*}

\begin{figure*}
\centering
\includegraphics[width=15.0cm]{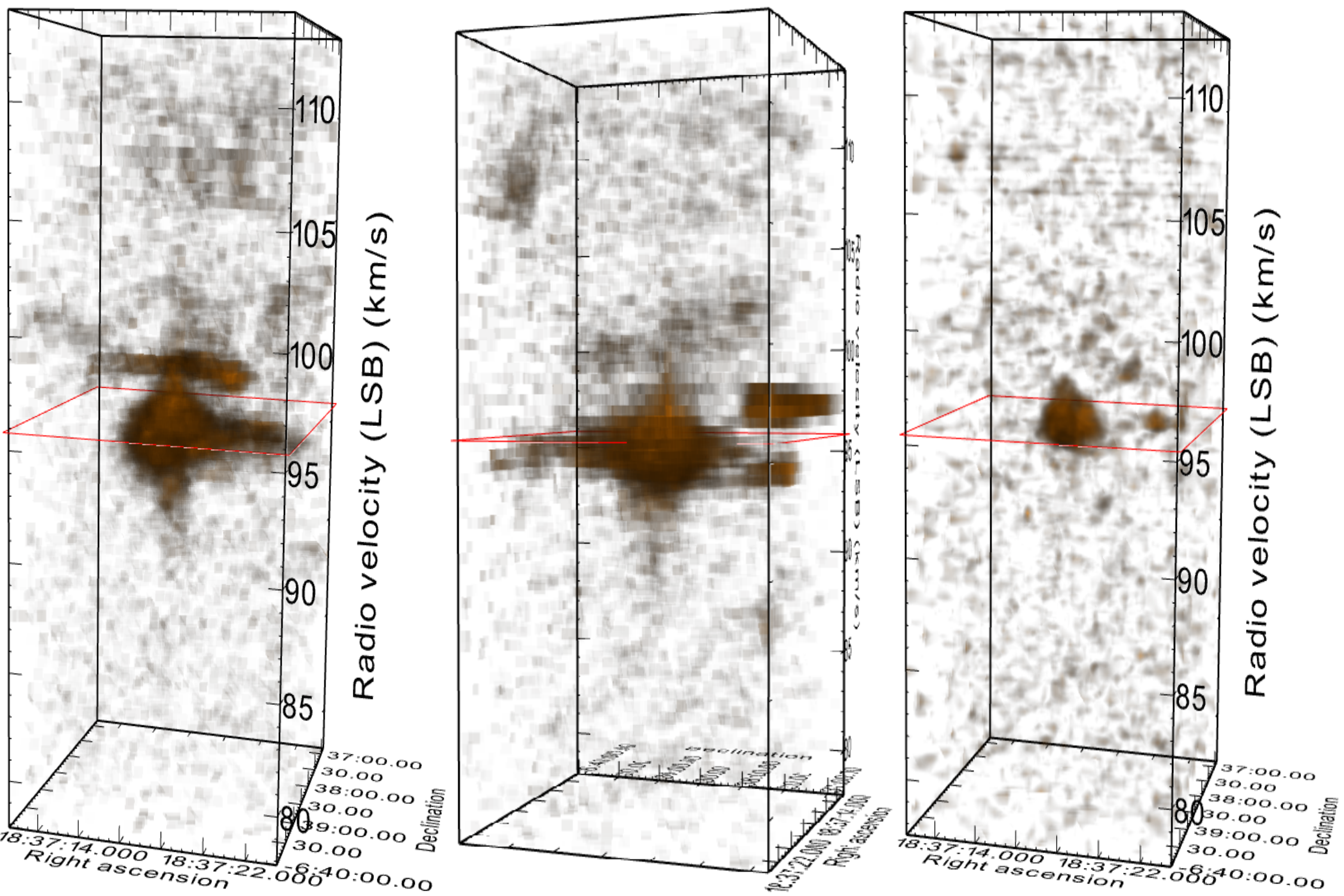}
\caption{$^{13}$CO~(J=3-2) and C$^{18}$O~(J=3-2) spectral 
cubes with the same spatial and spectral coverages as
in Figure~\ref{12CO-cube}.  The left and central  panels
show the $^{13}$CO cube from two different orientations
and the right panel shows the C$^{18}$O cube.
The red square shows the plane of the
$\upsilon$\textsubscript{LSR} of the core.}
\label{13CO-C18O-cube}
\end{figure*}

\begin{figure}
\centering
\includegraphics[width=9cm]{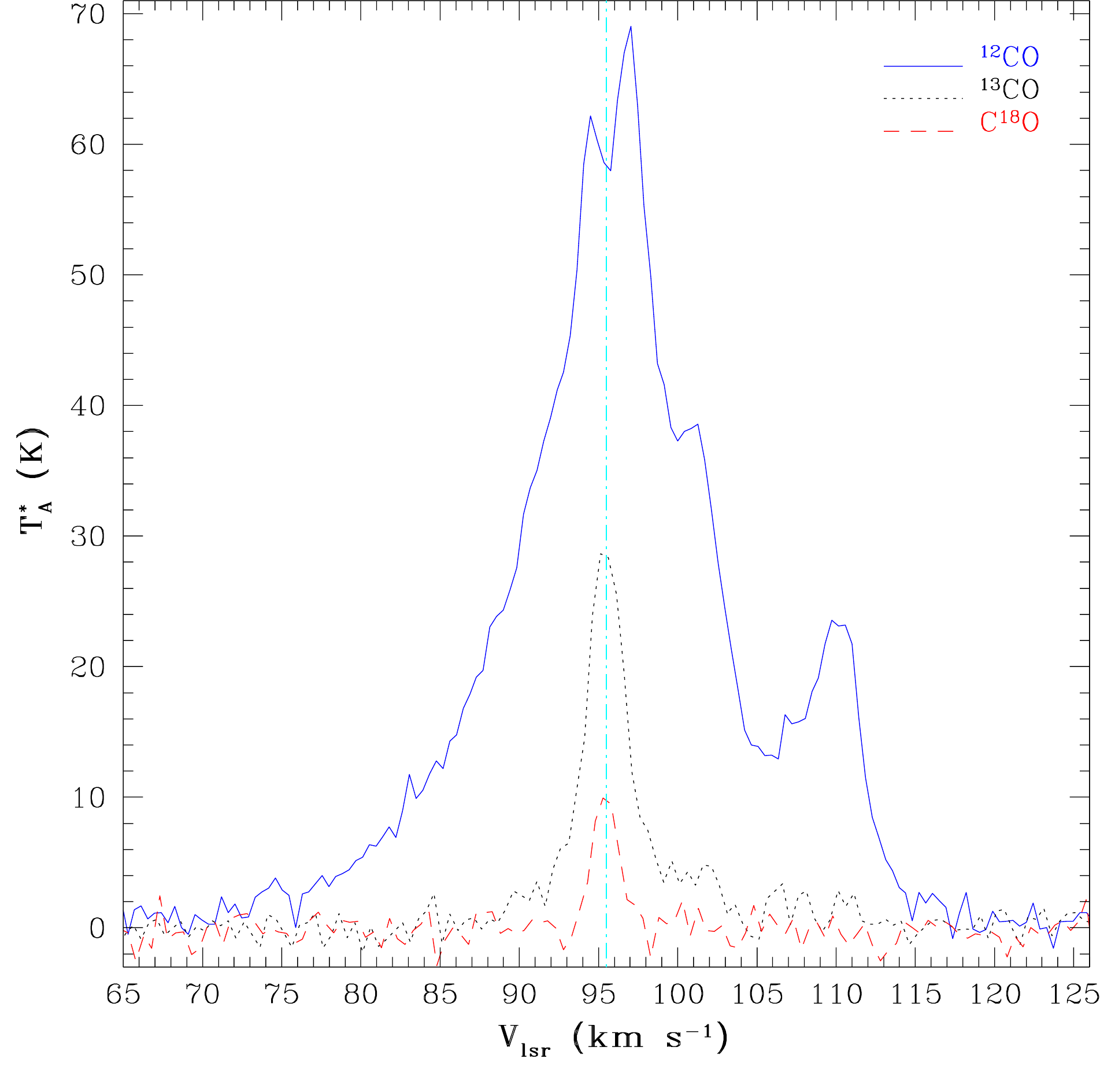}
\caption{The $^{12}$CO, $^{13}$CO and C$^{18}$O~(J=3-2) 
spectra integrated in a 1\arcmin$\times$1\arcmin-box
around IRAS~18345. The dotted-dashed vertical line at 95.5~km~s$^{-1}$ 
shows the {\bf $\upsilon$\textsubscript{LSR}} of the core.
The emission at 110~km~s$^{-1}$ is from the
ambient medium.}
\label{COspectra}%
\end{figure}

\section{Results and discussion}

\subsection {The outflows}

Through deeper observations in H$_2$ and $K$, we 
confirm the tentative detection of an H$_2$ outflow by
Varricatt et al. (\cite{varricatt10}).  The continuum
subtracted H$_2$ image in Figure~\ref{wfcamJHKH2} shows 
the emission features labelled MHO~2211, which is composed 
of two knots directed away from `B'.  We derive an outflow 
angle of 95$^{\circ}$ east of north if these are produced by 
an outflow from a source at the location of `B'.

High angular resolution VLBI observations of the 6.7\,GHz 
Class-II methanol maser by 
Bartkiewicz et al. (\cite{bartkiewicz09}) revealed 30 maser 
spots distributed in the form of an ellipse with semi-axes 
103\,mas and 70\,mas respectively. They conclude that 
the 6.7\,GHz methanol masers form in a disk or torus around 
the massive YSOs, at the interface between the disk and the 
outflow, and that these masers form in a stage before the 
formation of the UCH{\sc{ii}} region. They estimate a radius 
of 980~AU for the ring of distribution of the maser spots, 
for a distance of 9.5\,kpc.  The coordinates of the brightest 
maser spot given by Bartkiewicz et al. 
($\alpha$=18:37:16.92106 $\delta$=-06:38:30.5017) agrees with 
those of the infrared source `B2' within 0.25\arcsec. It is 
also noteworthy that the major axis of the ring-like 
distribution of the maser spots is at an angle of 90$^\circ$, 
which is nearly perpendicular to the direction of orientation 
of the CO outflow, as it would be expected if they form in a 
disk or torus around the YSO.  Assuming a circular distribution 
for the maser spots, we infer an outflow angle of 
43$^{\circ}$ with respect to the sky plane.

The CO\,(3--2) mapping using JCMT (\S\ref{COdata})
reveal a well-defined outflow (Figures \ref{12CO-cube}, 
\ref{COspectra}). Contours generated from the blue- and 
red-shifted lobes of the outflow are overlaid on the H$_2$ 
image in Figure~\ref{wfcamJHKH2}.  The contours start 
at 3\,$\sigma$ above the average background level in the 
integrated blue and red lobes of the outflow.
The outflow is oriented approximately N-S; we measure 
an angle of 161$^{\circ}$ E of N for the blue lobe.  
The outflow appears to be centred on source `B'
and is in good agreement with the outflow from this
source detected in $^{12}$CO~(2--1) by 
Beuther et al. (\cite{beuther02b}). Figure \ref{wfcamJHKH2} 
shows that the direction of this 
outflow detected in CO is very different when 
compared with an angle of 95$^{\circ}$ derived for the H$_2$ outflow.
The outflow appears to be at a high inclination with 
respect to the sky plane, as inferred from the overlapping 
blue- and red-shifted lobes (Figure~\ref{wfcamJHKH2}) 
and the high velocity range (Figure~\ref{COspectra});
we adopt the angle of inclination of 43$^{\circ}$ with
respect to the sky plane inferred from the  
methanol masers for the calculations in this paper.

The CO spectral line data are used to determine the 
outflow properties. The velocity-integrated blue- 
and red-shifted spectral images were deconvolved using 
a 2D Gaussian of FWHM=14.58\arcsec, which is the beam 
size of the JCMT at this frequency. At 3-$\sigma$ 
level on the blue- and red-shifted images, 
the projected outflow length is 28\arcsec, and the 
width for the blue-shifted lobe is 47\arcsec. After 
correcting the observed length for an angle of 
inclination of 43$^{\circ}$, we derive a collimation 
factor of 0.81 for the outflow. 
Using a distance of 9.5\,kpc, and an average 
of the maximum observed velocities of the blue and 
red-shifted lobes of the outflow of 18~km~s$^{-1}$,
we derive the outflow length and dynamical time as 
1.76\,pc and 9.58$\times$10$^4$~years
respectively.

The optical depth in $^{12}$CO ($\tau_{12}$) can be obtained
from a ratio of the antennae temperatures (T$^*_A$) obtained 
in $^{12}$CO and $^{13}$CO assuming $^{12}$CO/$^{13}$CO=89,
and using Eq. \ref{eq_tau}  (Garden et al. \cite{garden91}).

\begin{equation}
\label{eq_tau}
\frac{T^*_A(^{12}CO)}{T^*_A(^{13}CO)} = \frac{1-e^{-\tau_{12}}}{1-e^{-\tau_{13}}} = \frac{1-e^{-\tau_{12}}}{1-e^{-\frac{\tau_{12}}{89}}}
\end{equation}

Since $^{12}$CO suffers from self absorption near the 
central velocity of the line (Figure \ref{COspectra}), 
we estimated $\tau$ separately in the blue and red wings 
adopting velocity ranges where both $^{12}$CO and $^{13}$CO 
were detected. We derive values of 18.85 and 15.65 for 
$\tau_{12}$ in the blue and red wings respectively.  

The $^{12}$CO spectrum exhibits a strong emission component 
at $\sim$110\,km~s$^{-1}$ (Figures \ref{12CO-cube} and 
\ref{COspectra}), which is probably not related to the 
outflow, and is arising from the ambient medium. This 
feature was interpolated out from the spectrum before 
estimating the emission in $^{12}$CO from the blue- and 
red-shifted lobes of the outflow. We used velocity ranges
77--90~km~s$^{-1}$ and 102--113~km~s$^{-1}$ respectively
for integrating the line wing emission in the blue-
and red-shifted lobes.

Following Garden et al. (\cite{garden91}), the $^{12}$CO
column density is derived using the equation

\begin{equation}
N = \frac{3k}{8\pi^3B\mu^2} \frac{e^\frac{hBJ(J+1)}{kT\textsubscript{ex}}}{(J+1)} \frac{(T\textsubscript{ex} + \frac{hB}{3k})}{[1-e^\frac{-h\nu}{kT\textsubscript{ex}}]} \int \tau_{\upsilon}~d\upsilon~cm^{-1},
\end{equation}
where {\em B} is the rotational constant, $\mu$ is the electric dipole moment
of the molecule, $J$ is the lower level of the transition, and $T$\textsubscript{ex}
is the excitation temperature.  Substituting
the values for {\em B} and $\mu$ in Eq. 2,
\begin{equation}
N = 2.39\times10^{14} \frac{e^\frac{hBJ(J+1)}{kT\textsubscript{ex}}}{(J+1)} \frac{(T\textsubscript{ex} + 0.92)}{[1-e^\frac{-16.6}{T\textsubscript{ex}}]} \int \tau_{\upsilon}~d\upsilon~cm^{-1},
\end{equation}
where $\upsilon$ is in km~s$^{-1}$.

Allowing for beam filling factor and ignoring the effect of 
the cosmic background radiation,
\begin{equation}
N = \frac{2.39\times10^{14}}{16.6} \frac{e^{\frac{hBJ(J+1)}{kT\textsubscript{ex}}}}{(J+1)} \frac{(T\textsubscript{ex} + 0.92)}{e^{\frac{-16.6}{T\textsubscript{ex}}}}
\int \frac{T^*_A}{\eta_b} \frac{\tau_{\upsilon}~d\upsilon}{(1-e^{-\tau_{\upsilon}})}~cm^{-1}.
\end{equation}

Adopting 30\,K for $T$\textsubscript{ex} 
(Beuther et al. \cite{beuther02b}) and $10^4$ for H$_2$/CO, Eq. 4 yields
1.37$\times$10$^{21}$~cm$^{-2}$ and 1.03$\times$10$^{21}$~cm$^{-2}$
respectively for the column density of H$_2$ in the blue- and 
red-shifted lobes.

The mass in the outflow lobes can be obtained using the 
relation $m~\times~nH_2~\times~S$, where $m$ is the mean 
atomic weight, which is 1.36$\times$mass of the hydrogen 
molecule, and $S$ is the surface area of the outflow lobe 
(Garden et al. \cite{garden91}).  At 3$\sigma$ above 
the mean background, we measure projected sizes (major 
axis $\times$ minor axis) of 55\arcsec$\times$45\arcsec 
and 30\arcsec$\times$22\arcsec respectively for the 
blue- and red-shifted lobes of the outflow. Integrating
within these regions, within the velocity ranges of
77--90~km~s$^{-1}$ and 102--113~km~s$^{-1}$ respectively,
we get 123\,M$_\odot$ and 25\,M$_\odot$ for the masses 
contained in the blue- and red-shifted lobes of the outflow. 
This gives a total mass of 148\,M$_\odot$ for the material 
in the outflow. Using the maximum observed velocities
($\upsilon$\textsubscript{max}) 18.5~km~s$^{-1}$ and 17.5~km~s$^{-1}$ 
respectively for the blue and red wings, and ignoring 
the angle of inclination of the outflow (see Cabrit \& 
Bertout	\cite{cabrit92}), we derive an outflow momentum
of 2662~M$_{\odot}$~km~s$^{-1}$, energy of 
45$\times$10$^{46}$~ergs, and a mass entrainment rate 
($\dot{M}$\textsubscript{out}) of 15.6$\times$10$^{-4}$~M$_{\odot}$~year$^{-1}$.
The total mass in the outflow estimated by us is very similar 
to 143\,M$_\odot$ estimated by Beuther et al. (\cite{beuther02b})
from CO\,(2--1).
However, the outflow momentum, energy and mass entrainment
rate estimated by us are higher. The difference comes from
the higher value of $\upsilon$\textsubscript{max} used by us, and the 
outflow momentum, energy and mass entrainment rate estimated 
here should be treated as upper limits only.

The collimation factor $f_c$=0.81 we obtain from 
the CO\,(3--2) map is lower than a value of 1.5 derived by 
Beuther et al. (\cite{beuther02b}) from their CO\,(2--1) 
map observed at a higher spatial resolution. Note that 
the deconvolved image of the blue-shifted lobe is highly 
asymmetric (see the blue contours in 
Figure \ref{wfcamJHKH2}). If this asymmetry is due to 
two overlapping outflows, the collimation factor will 
be higher; using a width of 26\arcsec of the 
red-shifted outflow lobe, we get a collimation factor 
of 1.46. With the direction of the outflow derived from 
the aligned H$_2$ knots much different 
from the direction of the main outflow mapped in CO, and
roughly agreeing with the direction of the asymmetry of
the blue lobe of the CO outflow as seen in Figure \ref{wfcamJHKH2},  
it is very likely that this region hosts at least two YSOs
in outflow phase.

\subsection {The sources driving the outflows}

Figure~\ref{wfcamJHKH2} shows that the CO outflow is centred
very close to the near-IR source `B'.  The H$_2$ line emission
knots (MHO~2211) also appear to be tracing back to the location 
of `B'. Figure~\ref{wfcamJHKcol} shows a ($J-H$) - ($H-K$) 
colour-colour diagram constructed using the objects detected in
a 2\arcmin$\times$2\arcmin field around IRAS~18345.  Since the 
UKIDSS observations were performed in better sky conditions 
than our current observations, UKIDSS data are used to construct 
the colour-colour diagram.  For `A' and `B', the most recent
magnitudes from Table \ref{tab:wfcamJHK} are used.  `B' for 
which $H$ and $J$ magnitudes are upper limits are shown 
with upward and rightward directed arrows to show the directions 
in which it move in the diagram with reliable detection in these 
bands. Both objects are at very large extinction. Source `B'
exhibits large excess and is located in the region of the 
colour-colour diagram occupied by luminous YSOs. The $K$-magnitudes 
given in Table \ref{tab:wfcamJHK}, observed from 2003 to 2012 
(all in the same photometric system and using the same aperture), 
show that `B' is a variable. 

The {\it Spitzer}-IRAC colour-colour diagram is useful 
to identify YSOs. Figure~\ref{IRAC-col} shows the colours 
of objects detected in the IRAC images, within a 
10\arcmin$\times$10\arcmin field around IRAS 18345. The dashed 
horizontal line shows the boundary between Class I/II (below) 
and Class I objects (above), and the dotted red arrows show
reddening vectors for A$_V$ = 50 derived using the extinction
law given in Mathis (1990). 
Objects with ([3.6]-[4.5])$>$0.8 and ([5.8]-[8.0])$>$1.1 
are likely to be Class I objects, which are protostars with 
infalling envelopes. Those with 0$<$([3.6]-[4.5])$<$0.8
and 0.4$<$([5.8]-[8.0])$<$1.1, within the region shown by 
the rectangle, are likely to be Class II objects, which are
stars with discs.  Most of the objects are located around 
([5.8]-[8.0], [3.6]-[4.5]) = 0,
which are foreground and background stars and discless 
pre-main-sequence (Class III) objects (Allen et al. \cite{allen04}; 
Megeath et al. \cite{megeath04}; Qiu et al. \cite{qiu08}).
Locations of `A' and `B' are shown by filled red-circles
in Figure~\ref{IRAC-col}. Source ``A'' is located in
the region occupied by young sources with discs and large 
reddening, and `B' is located in the region of even younger 
sources at larger reddening and an infalling envelopes. 
However, note that the locations 
of these objects could be affected by saturation, especially
in the 8$\mu$m band. The locations of the two sources in 
Figures \ref{wfcamJHKcol} and \ref{IRAC-col} suggest
that `B' is younger than `A'.

All data available with good spatial resolution (UKIRT-$J,H,K,L',M'$ 
and 12.5\,$\mu$m and {\it Spitzer}-IRAC) data show that source `B' 
has a steeper SED than `A'.  At 12.5\,$\mu$m, `B' contributes
25.6\% of the total flux of `A' and `B' combined, whereas in
$L'$ (3.75\,$\mu$m) and $M'$ (4.7\,$\mu$m), `B' contributes only 
4.3\%  and 12.4\% respectively of total flux of `A' and `B' 
combined.  At shorter wavelengths (UKIRT, {\it Spitzer}-IRAC and
WISE 3.4\,$\mu$m), `A' is the leading source. The longer
wavelength observations ({\it Spitzer}-MIPS, WISE-24\,$\mu$m,
JCMT and IRAM) do not have the spatial resolution to
resolve `A' and `B'.
However, the excellent positional accuracies
of these observations enable us to identify the
leading YSO with good confidence.
Figure \ref{IRAC123} shows a colour composite image
of a 1\arcmin$\times$1\arcmin field surrounding IRAS~18345,
constructed from {\it Spitzer}-IRAC images in bands 1, 2 and 3.
Locations of sources detected in different observations
with good positional accuracies are overlaid on the
image. As can be seen in Figure \ref{IRAC123} all longer
wavelength observations peak near source `B'.  These 
factors leads us to infer that `B (B1+B2)' or some embedded 
sources very close to it are the YSOs responsible 
for the outflows in this system.

\begin{figure}
\centering
\includegraphics[width=8.9cm]{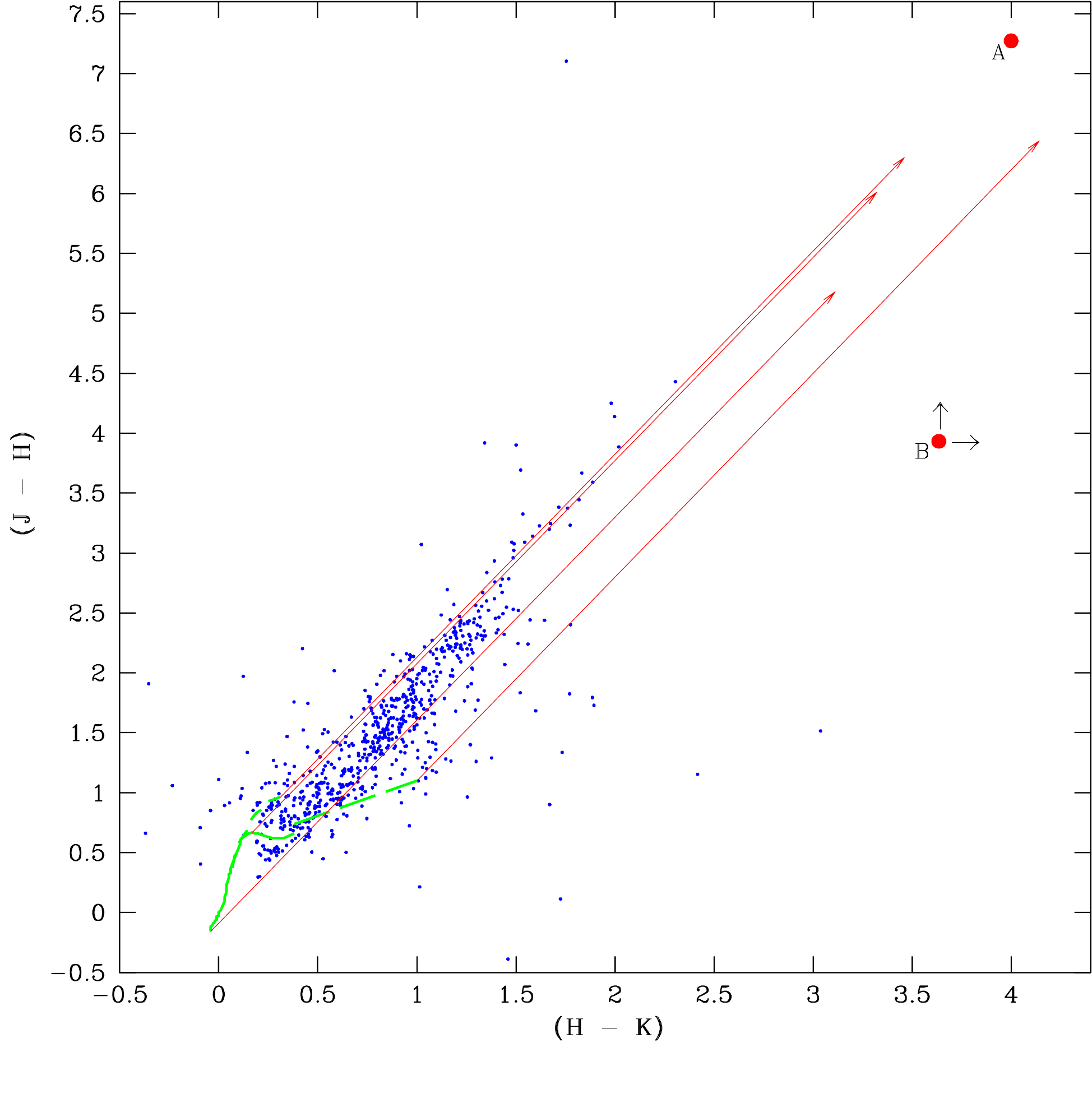}
\caption{WFCAM $JHK$ colour-colour diagram
of a 2\arcmin$\times$2\arcmin region surrounding IRAS~18345;
UKIDSS data obtained in better sky conditions are used.
The blue dots show the colours of point sources detected.
The red dots are the colours of
sources `A' and `B' derived from the most recent observations.
The continuous green lines show the colours of unreddened
main sequence and red-giant stars taken from 
Tokunaga et al. (\cite{tokunaga00}). The long-dashed green 
line shows the loci of CTTS from Meyer et al. (\cite{meyer97}).
The long red arrows show the reddening vectors for A$_V$ $\leq$ 50 
derived from the interstellar extinction law given by Rieke \&
Lebofsky (1985).
}
\label{wfcamJHKcol}%
\end{figure}

\begin{figure}
\centering
\includegraphics[width=8.9cm]{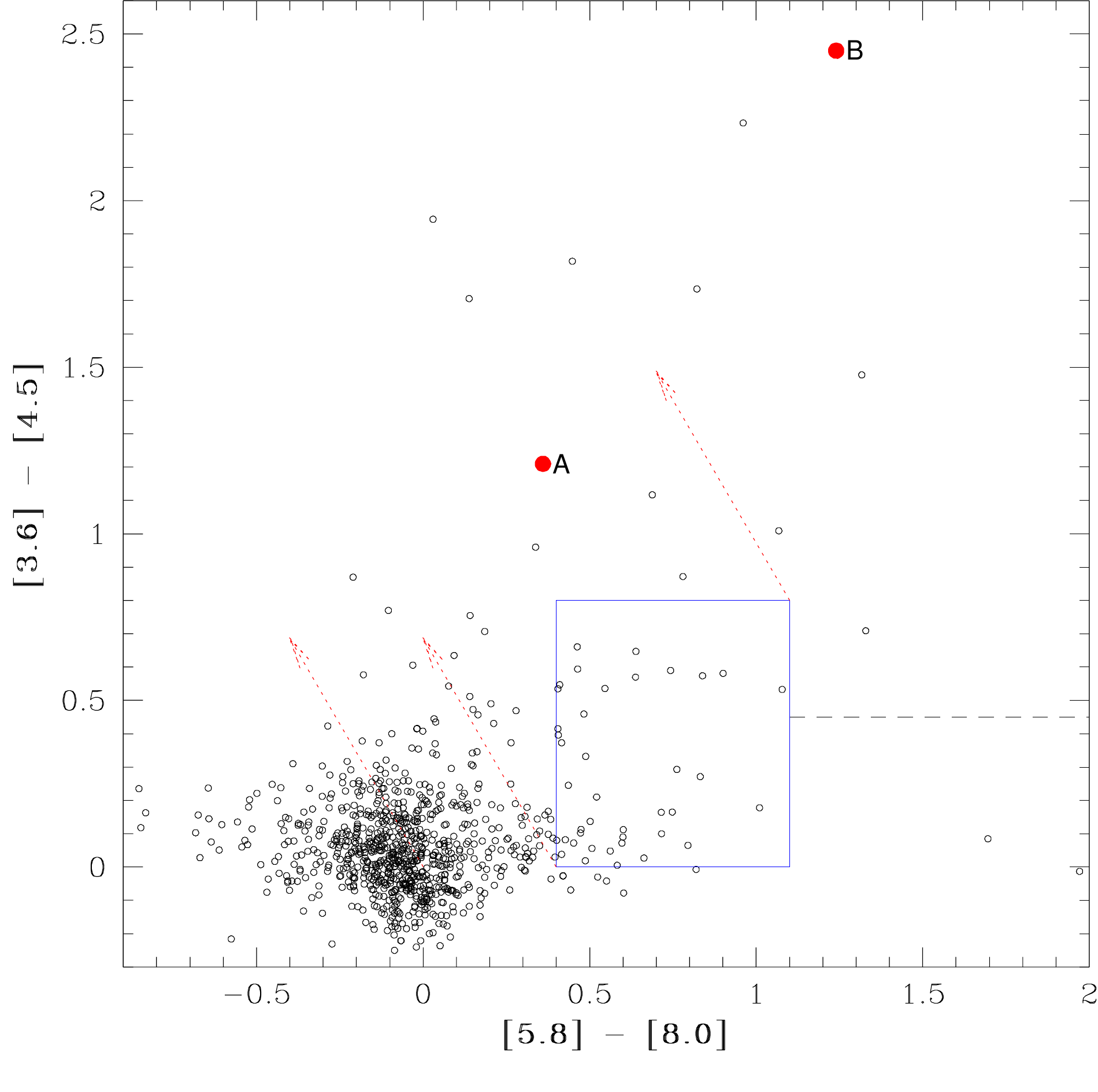}
\caption{The IRAC colour-colour diagram.  The open 
(black) circles show the objects detected in all 
four IRAC bands in a 10$\arcmin\times$10$\arcmin$
field centred on IRAS~18345.  The locations of
sources `A'' and ``B'' are shown by filled (red) 
circles.  The loci of Class II sources are shown
by the blue rectangle;  the dashed horizontal line 
shows the boundary between Class I/II objects (below) 
and Class I objects (above).  
The dotted red arrows show reddening vectors for
A$_V$=50.
}
\label{IRAC-col}%
\end{figure}

\subsubsection{The spectral energy distribution}
\label{fitting}

We modelled the SED of IRAS~18345 using the SED fitting tool 
of Robitaille et al. (\cite{robitaille07}), which uses a grid 
of 2D radiative transfer models presented in 
Robitaille et al. (\cite{robitaille06}), developed by 
Whitney et al. (\cite{whitney03a}, \cite{whitney03b}, etc).  
The grid consists of SEDs of 20000 YSO models
covering a range of stellar masses from 0.1 to 50~M$_\odot$ and
and evolutionary stages from the early envelope infall stage to
the late disc-only stage, each at 10 different viewing angles.
With all the longer wavelength data peaking close to `B',
we have treated that as the leading YSO in this region.

Up to 12.5\,$\mu$m, we have fluxes of `B' available.
WFCAM $K$, IRCAM $L' M'$, {\it Spitzer}-IRAC Bands 1 and 3
and MICHELLE 12.5-$\mu$m flux of `B' were used.  WFCAM $H$
magnitude was not used since in $H$, we are probably seeing 
only the nebulosity associated with `B'. {\it Spitzer}-IRAC 
Bands 2 and 4 fluxes were not used since they are near
saturation.  The resolution of WISE images is not sufficient
to resolve the two sources, so WISE 3.4--12-$\mu$m data 
{\bf were} not used to construct the SED.  At 22\,$\mu$m, the 
WISE source is located close to `B', there is still 
likely to be a significant contribution to the flux from 
`A'.  Hence the 22-$\mu$m flux was used as an upper 
limit. IRAS catalog lists the flux values for IRAS~18345 
in 12 and 100\,$\mu$m bands as lower limits, and the 
100-$\mu$m flux may be affected by infrared cirrus, so those 
values were not used in the SED analysis.  AKARI-IRC 
9-$\mu$m flux was not used
since that wavelength was already covered by the higher
resolution UKIRT data.  Similarly, MSX data in Bands A, C 
and D were not used.  AKARI-IRC 18-$\mu$m data and MSX
Band-E data are treated as upper limits only since they
contain contributions from source `A'.  The large
bandwidths of the IRAS and AKARI photometry warrant 
colour corrections, which were performed using the 
correction factors given in their respective point 
source catalogs (Beichman et al. \cite{beichman88}; 
Kataza et al. \cite{kataza10}; Yamamura et al. \cite{yamamura10}). 
The JCMT-SCUBA 450 and 850-$\mu$m fluxes were adopted 
from Williams, Fuller \& Sridharan (\cite{williams04}).
Flux estimates at 1.1~mm and 1.2~mm were adopted from
Rosolowsky et al. (\cite{rosolowsky10}) and
Beuther et al. (\cite{beuther02a}) respectively.
Figure~\ref{18345-sed} shows the model fit to the spectral
energy distribution. 
Table \ref{tab:results} shows the results of the SED 
modelling.  The model fit to the SED yields a very young 
object of mass 19.8\,M$_\odot$, age 3.3$\times$10$^3$ years and
luminosity 1.9$\times$10$^4$\,L$\odot$.  
Figure~\ref{18345-sed} shows that there is a large
disagreement between the fluxes obtained by different
telescopes in the wavelength range of 60--140\,$\mu$m.
The flux estimated from the {\it Spitzer}-MIPS 70\,$\mu$m 
image is much lower than what was seen in IRAS 60\,$\mu$m
and AKARI 66\,$\mu$m bands.  Similarly, the AKARI 
90 and 140-$\mu$m fluxes also deviate significantly 
from the model SED.  Inadequate spatial 
resolution and the different contributions from the background
could be severely affecting the fluxes measured
in these bands used by telescopes of different
sizes and spatial resolutions. 
The SED analysis for this source available in the literature,
performed using the same model (Tanti, Roy \& Duorah, \cite{tanti11})
using all the available data, obtained a 12.99~M$_\odot$ YSO
with an age of 1.36$\times$10$^4$~years and a luminosity of
7.21$\times$10$^3$~L$_\odot$.  The difference
mostly comes from the fact that they do not attempt
to separate the fluxes of `A' and `B'.

The results obtained from our SED analysis bring some 
complications. The dynamical time scale of the outflow 
(9.58$\times$10$^4$~years) derived by us from the
CO\,(3--2) data is much larger
than the age  of the YSO (3.3$\times$10$^3$\,years) obtained 
from the SED analysis. This is unrealistic. This 
discrepancy may mostly be caused by the contribution from 
a companion at longer wavelengths, thereby making the star 
appear younger.  We have removed the contributions from 
source `A' to the total flux for wavelengths up to 
12.5\,$\mu$m.  The centroids of all detections at longer 
wavelengths are located close to `B', hence `A' is
not likely to be the main contributor at longer wavelengths.
The more probable chance is that `B' itself is a binary 
or multiple as strongly suggested by its resolution into 
two components `B1' and `B2' in $L'$ and $M'$, and the 
detection of outflows with two different directions from 
H$_2$ and CO observations, both of which appear to be
centred near `B'.  High-angular-resolution 
observations at longer wavelengths are therefore required 
to properly understand the complexity of the star formation 
taking place here.

\begin{figure}
\centering
\includegraphics[width=9.1cm]{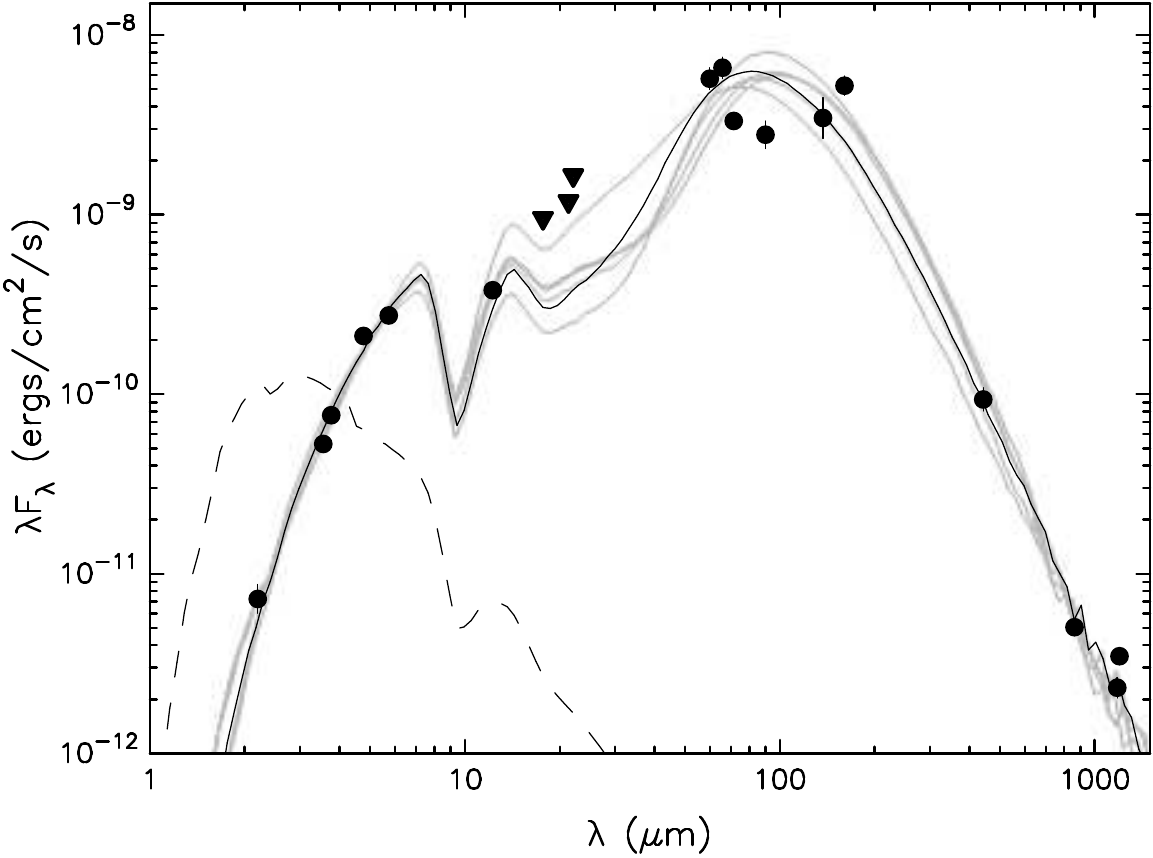}
\caption{The SED of IRAS~18345. The filled circles show the data 
in WFCAM $K$ band,
IRCAM $L'$ and $M'$ bands, {\it Spitzer}-IRAC Bands 1 and 3, 
Michelle 12.5\,$\mu$m, IRAS 60\,$\mu$m, AKARI at 65--140\,$\mu$m,
{\it Spitzer}-MIPS 70\,$\mu$m, SCUBA 450 and 850\,$\mu$m,
Bolocam 269 GHz and IRAM 240 GHz data.  The downward-directed
triangles show AKARI-IRC 18\,$\mu$m, MSX band-E and WISE
band-4 data which are treated as upper limits.  The continuous 
line shows the best fit model. The grey lines show subsequent 
good fits for ($\chi^2-\chi^2_{best fit}$) per data point $<$ 3.
The dashed line corresponds to the stellar photosphere for the
central source of the best fitting model, as it would look
in the absence of circumstellar dust (but including interstellar
extinction).}
\label{18345-sed}%
\end{figure}

\begin{table}
\caption{Results from the SED fitting}             
\label{tab:results}      	
\centering                      
\begin{tabular}{ll}        	
\hline\hline								\\[-2mm]
Parameter                       &Best-fit values$^{\mathrm{a}}$		\\
\hline              							\\[-2mm]         
Stellar mass (M$_{\odot}$)      &19.8 (+3, -1.5)			\\
Stellar age (yr)                &3.3 (+.49, -0.77)$\times$10$^{3}$	\\
Stellar radius (R$_{\odot})$	&230 (+16, -20)				\\
Stellar temperature (K)		&4480 (+260, -167)			\\
Disk mass (M$_{\odot}$)         &1.9 (+6, -0)$\times$10$^{-1}$		\\
Disk accretion rate (M$_{\odot}$yr$^{-1}$)  &2.5 (+4, -1.4)$\times$10$^{-5}$	\\
Disk/envelope inner radius (AU) &9.9 (+0, -1.3)				\\
Disk outer radius (AU)		&11.7 (+11.8, -0)			\\
Envelope mass (M$_{\odot}$)     &2.2 (+3.6, -0)$\times$10$^{3}$    	\\
Envelope accretion rate (M$_{\odot}$yr$^{-1}$)  &4.1 (+4.3, -0.2)$\times$10$^{-3}$ \\
Envelope outer radius (AU)	&1 (+0, -0.4)$\times$10$^{5}$		\\
Total Luminosity (L$_{\odot}$)  &1.9 (+2, -1.5)$\times$10$^{4}$		\\
A$_V$ circumstellar (mag)	&804 (+508, -453)			\\
A$_V$ interstellar (mag)   	&25 ($\pm$13)				\\
Angle of inclination of the disk axis ($^{\circ}$)	&18		\\
\hline 
\end{tabular}
\begin{list}{}{}
\item[$^{\mathrm{a}}$] The values given in parenthesis are the 
rms of the differences, in parameter values of models with 
($\chi^2-\chi^2_{best fit}$) per data point $<$ 3, above and 
below those of the best-fit model, estimated with respect to the 
parameter values of the best-fit model.  A distance of 9.5\,kpc 
is adopted.
\end{list}
\end{table}


\section{Conclusions}

\begin{enumerate}
\item IRAS~18345 is a very young massive YSO in a phase
of active accretion.
\item Through near and mid-IR observations, we identify
the central source, which is highly obscured and is 
probably double at a separation of $\sim$0.45\arcsec.
\item We confirm the 2.122\,$\mu$m H$_2$ outflow tentatively 
detected by Varricatt et al. (\cite{varricatt10}). 
The direction of this outflow is very different from the 
direction of the main outflow mapped in CO. The origins
of both H$_2$ and CO outflows appear to trace back to the location
of the infrared source mentioned above.  In addition, the SED 
analysis does not fit well with a single YSO model. This 
suggests that more than one YSO are driving outflows here.
\item The SED analysis yields a 19.8\,M$_\odot$ star of 
age 3.3\,$\times$10$^3$\,years.  The age is far lower than
what is implied by the dynamical time of the outflow.
The unresolved binarity is probably the factor affecting
the SED analysis, resulting in an overestimate of the 
mass and underestimate of the age.
\item The collimation factor of the outflow derived by us 
from the CO~(3-2) maps is low (0.81). It needs to be explored 
if this is real, or is due to the presence of more than one 
outflow.
\item We emphasize the need high-angular-resolution 
observations at mid-IR and longer wavelengths to properly 
understand this system, and in general massive star forming 
regions. High angular resolution interferometric CO line 
observations are also required to understand if the apparently 
single outflow seen in CO is composed of more than one component.
\end{enumerate}

\begin{acknowledgements}
The UKIRT is operated by the Joint Astronomy Centre (JAC) 
on behalf of the Science and Technology Facilities Council 
(STFC) of the UK. The JCMT is operated by the JAC on 
behalf of the STFC, the Netherlands Organization for 
Scientific Research, and the National Research Council 
of Canada. The UKIRT data presented in this paper are obtained 
during the UKIDSS back up time. We thank  Thor Wold, Tim Carroll 
and Jack Ehle for carrying out the WFCAM observations, 
the Cambridge Astronomical Survey Unit (CASU) for processing 
the WFCAM data, and the WFCAM Science Archive (WSA) for 
making the data available.  The
CO~(3-2) data are obtained from the JCMT science archive.  
This work makes use of data obtained with AKARI, a JAXA project 
with the participation of ESA, and IRAS data downloaded from 
the SIMBAD database operated by CDS, Strasbourg, France.
The archival data from {\it Spitzer}, WISE and MSX are downloaded 
from NASA/IPAC Infrared Science Archive, which is operated by the Jet Propulsion 
Laboratory, Caltec, under contract with NASA. We thank the anonymous
referee for the helpful comments and suggestions which have
improved the paper.  
\end{acknowledgements}


\begin{thebibliography}{}

\bibitem[2004]{allen04} Allen, L.~E. et al. 2004, ApJS, 154, 363
\bibitem[1990]{anderson90} Anderson, T., De Lucia, Fr, Herbst, E. 1990, ApJS, 72, 797
\bibitem[2007]{arce07} Arce, H.~G., Shepherd, D., Gueth, F., Lee, C.-F., Bachiller, R., Rosen, A., Beuther, H.  2007, Protostars and Planets V, 245 
\bibitem[1988]{beichman88} Beichman, C.~A.; Neugebauer, G., Habing, H.~J., Clegg, P.~E., Chester, T.~J. 1988, iras, 1
\bibitem[2009]{bartkiewicz09} Bartkiewicz, A., Szymczak, M., van Langevelde, H. J., Richards, A. M. S., Pihlstr\"{o}m, Y. M. 2009, A\&A, 502, 155
\bibitem[2002a]{beuther02a} Beuther, H., Schilke, P., Menten, K.~M., Motte, F., Sridharan, T.~K., Wyrowski, F. 2002a, ApJ, 566, 945
\bibitem[2002b]{beuther02b} Beuther, H., Schilke, P., Sridharan, T.~K., Menten, K.~M., Walmsley, C.~M., Wyrowski, F. 2002b, A\&A, 383, 892
\bibitem[2002c]{beuther02c} Beuther, H., Walsh, A., Schilke, P., Sridharan, T. K., Menten, K. M., Wyrowski, F.  2002c, A\&A, 390, 289
\bibitem[1996]{bronfman96} Bronfman, L., Nyman, L.~A., May, J. 1996, A\&AS, 115, 81
\bibitem[2009]{buckle09} Buckle, J. V., Hills, R. E., Smith, H., and 35 coauthors 2009, MNRAS, 399, 1026
\bibitem[1992]{cabrit92} Cabrit, S., Bertout, C. 1992, A\&A, 261, 274
\bibitem[2007]{casali07} Casali, M. et al. 2007, A\&A, 467, 777
\bibitem[2008]{cavanagh08} Cavanagh, B., Jenness, T., Economou, F., Currie, M.~J. 2008AN....329..295C
\bibitem[2008]{currie08} Currie M.~J., Draper P.~W., Berry D.~S., Jenness T., Cavanagh B., Economou F. 2008, in Argyle R.~W., Bunclark P.~ S., Lewis J.~R., eds., Astron. Soc. of the Pac. Conf. Series Vol. 394, Astronomical Data Analysis Software and Systems XVII. p. 650
\bibitem[2004]{davis04} Davis, C.~J., Varricatt, W.~P., Todd, S.~P., Ramsay Howat, S.~K., 2004, A\&A, 425, 981
\bibitem[2009]{doi09} Doi, Y., Etxaluze Azkonaga, M., White, G., et al.  2009, sitc.conf, p.4018
\bibitem[2007]{edris07} Edris, K.~A., Fuller, G.~A., Cohen, R.~J. 2007, A\&A, 465, 865
\bibitem[2004]{fazio04} Fazio, G.~G. et al. 2004, ApJS, 154, 10
\bibitem[1992]{felli92} Felli, M., Palagi, F., Tofani, G. 1992, A\&A, 255, 293
\bibitem[1991]{garden91} Garden, R. P., Hayashi, M., Hasegawa, T., Gatley, I., Kaifu, N. 1991, ApJ, 374, 540
\bibitem[1997]{glasse97} Glasse, A.~C., Atad-Ettedgui, E.~I., Harris, J.~W. 1997, SPIE, 2871, 1197 (Proc. SPIE Vol. 2871, p. 1197-1203, Optical Telescopes of Today and Tomorrow, Arne L. Ardeberg; Ed.)
\bibitem[2008]{hambly08} Hambly, N.~C.; Collins, R.~S.; Cross, N.~J.~G., and 14 co authors 2008, MNRAS, 384, 637
\bibitem[2006]{hewett06} Hewett, P.~C., Warren, S.~J., Leggett, S.~K., Hodgkin, S.~T. 2006, MNRAS, 367, 454
\bibitem[2009]{hodgkin09} Hodgkin, S.~T., Irwin, M.~J., Hewett, P.~C., Warren, S.~J. 2009, MNRAS, 394, 675
\bibitem[2011]{hofner11} Hofner, P., Kurtz, S., Ellingsen, S.~P., Menten, K.~M., Wyrowski, F., Araya, E.~D., Loinard, L., Rodr{\'{i}}guez, L.~F., Cesaroni, R. 2011, ApJ, 739, L17
\bibitem[2004]{irwin04} Irwin, M. J., Lewis, J., Hodgkin, S., et al. 2004, in Optimizing Scientific Return for Astronomy through Information Technologies, eds. P. J. Quinn \& A. Bridger, Proc. SPIE, 5493, 411 
\bibitem[2010]{ishihara10} Ishihara, D., Onaka, T., Kataza, H., and 30 co-authors 2010, A\&A, 514A, 1
\bibitem[2010]{kataza10} Kataza, H. et al. 2010, AKARI-IRC Point Source Catalogue Release note Version 1.0
\bibitem[2007]{lawrence07} Lawrence, A., Warren, S.~J., Almaini, O., and 19 co authors 2007, MNRAS, 379, 1599
\bibitem[2005]{maret05} Maret, S., Ceccarelli, C., Tielens, A.~G.~G.~M., Caux, E., Lefloch, B., Faure, A., Castets, A., Flower, D.~R. 2005, A\&A, 442, 527
\bibitem[1990]{mathis90} Mathis, J.~S. 1990, ARA\&A, 28, 37
\bibitem[2004]{megeath04} Megeath, S.~T. et al. 2004, ApJS, 154, 367
\bibitem[1997]{meyer97} Meyer, M.~R., Calvet, N., Hillenbrand, L.~A. 1997, AJ, 114, 288
\bibitem[2007]{murakami07} Murakami, H. et al. 2007, PASJ, 59, S369
\bibitem[2005]{pestalozzi05} Pestalozzi, M. R., Minier, V., Booth, R. S. 2005, A\&A, 432, 737
\bibitem[2000]{prestage00} Prestage, R.~M, Meyerdierks, H., Lightfoot, J.~F, Jenness, T., Tilanus, R.~P.~J, Padman R. 2000, {\sc specx} - A Millimetre Wave Spectral Reduction Package, CCLRC / Rutherford Appleton Laboratory, Particle Physics \& Astronomy Research Council, Starlink User Note 17.8
\bibitem[2008]{qiu08} Qiu, K. et al. 2008, ApJ, 685, 1005
\bibitem[2004]{rieke04} Rieke, G.~H. et al. 2004, ApJS, 154, 25
\bibitem[1985]{rieke85} Rieke, G.~H., Lebofsky, M.~J. 1985, ApJ, 288, 618
\bibitem[2006]{robitaille06} Robitaille, T.~P., Whitney, B.~A., Indebetouw, R., Wood, K., and Denzmore, P. 2006, ApJS, 167, 256
\bibitem[2007]{robitaille07} Robitaille, T.~P., Whitney, B.~A., Indebetouw, R., Wood, K. 2007, ApJS, 169, 328
\bibitem[2003]{rodgers03} Rodgers, S.~D., Charnley, S.~B. 2003, ApJ, 585, 355
\bibitem[2010]{rosolowsky10} Rosolowsky, E., Dunham, M.~K., Ginsburg, A., Bradley, E.~T., Aguirre, J., Bally, J., Battersby, C., Cyganowski, C., Dowell, D., Drosback, M., and 6 coauthors   2010, ApJS, 188, 123
\bibitem[2002]{sridharan02} Sridharan, T.~K., Beuther, H., Schilke, P., Menten, K.~M., Wyrowski, F.  2002, ApJ, 566, 931
\bibitem[2000]{szymczak00} Szymczak, M., Hrynek, G., Kus, A.~J. 2000, A\&AS, 143, 269
\bibitem[2000]{szymczak02} Szymczak, M., Kus, A. J., Hrynek, G., K\u{e}pa, A., Pazderski, E. 2002, A\&A, 392, 277
\bibitem[2011]{tanti11} Tanti, K.~K, Roy, J., Duorah, K. 2011, IJSER, Vol. 2, Issue 10
\bibitem[2008]{thomas08} Thomas, H. S., Fuller, G. A. 2008, A\&A, 479, 751
\bibitem[2000]{tokunaga00} Tokunaga, A. T. 2000, in Allen's Astrophysical Quantities, ed. A. N. Cox (Springer-Verlag), 143
\bibitem[1995]{vanderwalt95} van der Walt, D.~J., Gaylard, M.~J., MacLeod, G.~C.  1995, A\&AS, 110, 81
\bibitem[2010]{varricatt10} Varricatt, W.~P., Davis, C.~J., Ramsay, S., Todd, S.~P. 2010, MNRAS, 404, 661
\bibitem[2003a]{whitney03a} Whitney, B.~A., Wood, K., Bjorkman, J.~E., Cohen, M. 2003a, ApJ, 598, 1079
\bibitem[2003b]{whitney03b} Whitney, B.~A., Wood, K., Bjorkman, J.~E., Wolff, M.~J. 2003b, ApJ, 591, 1049
\bibitem[2004]{williams04} Williams, S.~J., Fuller, G.~A., Sridharan, T.~K. 2004, A\&A, 417, 115
\bibitem[2010]{wright10} Wright, E.~L., Eisenhardt, P.~R.~M., Mainzer, A.~K., et al. 2010, AJ, 140, 1868
\bibitem[2010]{yamamura10} Yamamura, I., Makiuti, S., Ikeda, N., Fukuda, Y, Oyabu, S, Koga, T., White, G. J. 2010, AKARI-FIS Bright Source Catalogue Release note Version 1.0

\end{thebibliography}
\end{document}